\documentclass[fleqn,12pt]{article}
\usepackage{amsfonts,epsfig,latexsym}

\newcommand{\Ga}{\alpha}
\newcommand{\Gb}{\beta}
\newcommand{\GD}{\Delta}
\newcommand{\Gd}{\delta}

\newcommand{\Geps}{\varepsilon}
\newcommand{\Gg}{\gamma}

\newcommand{\CA}{{\cal A}}

\newcommand{\CK}{{\cal K}}
\newcommand{\CL}{{\cal L}}
\newcommand{\CP}{{\cal P}}

\newcommand{\CO}{{\cal O}}

\newcommand{\CU}{{\cal U}}
\newcommand{\CV}{{\cal V}}

\newcommand{\Bchi}{\overline{\chi}}

\newcommand{\zb}{{\bar{z}}}
\newcommand{\pb}{{\bar{p}}}

\newcommand{\ft}[2]{{\textstyle {\frac{#1}{#2}} }}

\newcommand{\dd}{\partial}
\newcommand{\tr}{{\rm tr \,}}
\newcommand{\Tr}{{\rm Tr \,}}

\newcommand{\ra}{\rightarrow}

\newcommand{\be}{\begin{equation}}
\newcommand{\ee}{\end{equation}}
\newcommand{\ben}{\begin{displaymath}}
\newcommand{\een}{\end{displaymath}}
\newcommand{\ba}{\begin{eqnarray}}
\newcommand{\ea}{\end{eqnarray}}
\newcommand{\nn}{\nonumber}
\newcommand{\non}{\nonumber\\}
\newcommand{\bean}{\begin{eqnarray*}}
\newcommand{\eean}{\end{eqnarray*}}

\newcommand{\mathon}{\mathversion{bold}}
\newcommand{\mathoff}{\mathversion{normal}}

\newcommand{\la}{\label}

\newcommand{\Ref}[1]{(\ref{#1})}

\makeatletter
\@addtoreset{equation}{section}
\makeatother

\newcommand{\cro}{\!\times\!}
\newcommand{\equ}{\!=\!}
\newcommand{\pls}{\!+\!}
\newcommand{\mis}{\!-\!}
\newcommand{\vl}{{\vphantom{[}}}

\newcommand{\Si}{{\bf i}}

\newcommand{\Mat}{{\mathcal S}}

\newcommand{\trgo}{{\rm Tr} \, g_{(1)}}
\newcommand{\trgt}{{\rm Tr} \, g_{(2)}}
\newcommand{\trh}{{\rm Tr} \, h_{(2)}}
\providecommand{\square}{\Box}
\newcommand{\Ksquare}{\square \hspace{-1.5ex}%
                       \raisebox{0.4ex}{{\it {\tiny K}}}\hspace{0.4ex} }
\newcommand{\zh}{\zeta}

\begin{document}

\thispagestyle{empty}

\begin{center}
IHES/P/02/70 \hspace*{1cm}
ROM2F/2002/23 \hspace*{1cm}
ITF-2002/46\hspace*{1cm}
SPIN-2002/28
\end{center}
\vspace*{1.2cm}

%\renewcommand{\thefootnote}{\fnsymbol{footnote}}

%\vspace*{0.05cm}
%\noindent
%\rule{\linewidth}{0.6mm}
%\vspace*{0.6cm}
\begin{center}
{\bf\LARGE Holographic Correlators \\ [5mm]
in a Flow to a Fixed Point}
\bigskip\bigskip

{\bf Marcus~Berg\footnotemark \,\, and
Henning~Samtleben\footnotemark}  

\vspace{.3cm}  
$^1${\em Department of Physics,
University of Rome, Tor Vergata \\
Via della Ricerca Scientifica, 00133 Rome, Italy\\
E-mail: {\tt berg@roma2.infn.it}}

\vspace{.5cm}

$^2${ {\em Institute for Theoretical Physics}  \& 
{\em Spinoza Institute,\\ 
Utrecht University, Postbus 80.195, 3508 TD Utrecht, 
The Netherlands\\
E-mail: {\tt h.samtleben@phys.uu.nl}}}

\end{center}
\renewcommand{\thefootnote}{\arabic{footnote}}
\setcounter{footnote}{0}
\bigskip
\medskip
\begin{abstract}
Using holographic renormalization, we study correlation functions
throughout a renormalization group flow between two-dimensional
superconformal field theories. The ultraviolet theory 
is an $N=(4,4)$ CFT which
can be thought of as a symmetric product of $U(2)$ super WZW
models. It is perturbed by a relevant operator
which preserves one-quarter supersymmetry and drives the theory to an
infrared fixed point.  We compute correlators of the stress-energy
tensor and of the relevant operators dual to supergravity
scalars. Using the former, we put together Zamolodchikov's
$C$~function, and contrast it with proposals for a holographic
$C$~function.  In passing, we address and resolve two puzzles also
found in the case of five-dimensional bulk supergravity.

\end{abstract}

\renewcommand{\thefootnote}{\arabic{footnote}}
\vfill
\leftline{{ September 2002}}

\setcounter{footnote}{0}

\newpage 
\setcounter{page}{1}
\renewcommand{\thepage}{\roman{page}}

{\baselineskip 13pt \tableofcontents}

\newpage
\renewcommand{\thepage}{\arabic{page}}
\setcounter{page}{1}

\section{Introduction}

In the AdS/CFT correspondence, renormalization group (RG) flows of a
$d$-dimensional conformal field theory are described by domain wall
solutions of the dual $(d+1)$-dimen\-sional bulk supergravity theory,
see e.g.~\cite{DHoFre02} and references therein.  Physically, the
domain wall solution can be thought of as a shell of matter, where the
metric becomes asymptotically AdS far from the shell.  In terms of
limiting behavior of the dual field theory, the asymptotic AdS length
scale gives the central charge at the ultraviolet conformal fixed
point of the field theory. Should the field theory be conformal also
in the infrared, the bulk space is asymptotically AdS also in the deep
interior of the shell, with the inner AdS length related to the
infrared central charge of the field theory.  In the supergravity
scalar target space this means that the flow does not run off to
infinity but goes down to a minimum of the potential and stops. The
solution is nonsingular everywhere, so for weak fluctuations and small
curvature, the supergravity approximation may be trusted throughout
the flow.

The previously studied five-dimensional examples of such
conformal-to-conformal flows could only be given
numerically~\cite{FGPW99}. This has remained an unsurmountable
obstacle for further application of holographic renormalization
methods, in particular for the computation of correlators in
nonsingular flows. The few flows that {\it are} known exactly, on the
other hand, all run off to infinity in the scalar target space, which
creates singularities at finite distance in the bulk
spacetime~\cite{BraSfe99,GPPZ00}. On the field theory side, these
solutions describe theories that confine in the infrared. In
\cite{BerSam01}, for three-dimensional bulk, we found the presently
only known exact solution describing an RG
 flow to an infrared fixed
point.  The present paper is about correlators in this smooth RG flow
between conformal fixed points. Our main result is the computation of
two-point functions of the stress-energy tensor and operators dual to
the supergravity scalars.\footnote{Guidelines for selected reading of
this paper are given at the end of the introduction.}

To be specific, the flow we study is a solution of the
three-dimensional $SO(4) \times SO(4)$ gauged supergravity
\cite{NicSam01b} with equal coupling constants and 16 supercharges.
The dual field theory is a large (or ``double'') $N=(4,4)$
superconformal theory with $SU(2)^4$ current algebra and equal levels
of the two $SU(2)^2$ factors.  It has an alternative realization as a
symmetric product of $U(2)$, $N=1$ super WZW models~\cite{dBPaSk99},
and in terms of branes, it is the worldvolume theory on the
intersection (along the D1-branes) of two D1-D5 systems with equal
D5-brane charge.  This brane setup has near-horizon limit $AdS_3
\times S^3 \times S^3 \times S^1$, where one can recognize the two
$SO(4)$'s as transverse rotations of the two brane systems.  Finally,
the RG flow we describe is driven by a relevant operator $\CO_q$ of
ultraviolet dimension $3/2$, that breaks conformal symmetry and leaves
1/4 of the supersymmetry.

Let us briefly expand on our motivations for this work.  On a purely
methodological level, we want to explore the extent to which
three-dimensional gauged supergravity can be used to describe
deformations of interesting but complicated two-dimensional CFTs,
such as worldvolume theories on intersecting D-branes.  In the end, we
hope to have provided an example that three-dimensional gauged
supergravity does provide a powerful addition to the arsenal of
methods in two-dimensional CFT.  In particular, on the supergravity
side we are able to compute quantities that are very difficult to
compute in the deformed CFT, e.g.\ $\langle TT \rangle$ correlators
and the $C$ function.  It is intriguing that on the field theory side,
those computations seem to be difficult but not impossible, which
makes our results predictions for supersymmetric (and perhaps
integrable) deformations of two-dimensional CFT.  In addition to
this motivation, our flow may be viewed as a toy model in which to
address some questions raised in flows of higher dimensional theories,
such as four-dimensional $N=1$ super Yang-Mills theory.  The toy model
idea proves to be useful, since we do find some new insight into old
mysteries encountered in holography with five-dimensional bulk
supergravity (see section~\ref{sec:twopoint}).

For completeness, let us mention that in \cite{BerSam01} we did not
just find the supersymmetric flow but we also found a stable
nonsupersymmetric fixed point of this theory (in fact, the first
stable nonsupersymmetric fixed point; later on several others were
found in the three-dimensional maximal theory
\cite{NicSam00,FiNiSa02}).  A flow to such a nonsupersymmetric fixed
point may be relevant to some aspects of black-hole physics, and this
was part of the motivation of~\cite{BerSam01}. However, although we
hope to discuss black-hole physics in future work, here we focus
entirely on the supersymmetric flow and the complementary motivations
mentioned in the previous paragraph.

To obtain the correlators, we use the formalism of holographic
renormalization \cite{dHSoSk00,BiFrSk01,BiFrSk01a,Sken02}. It is a
framework to reliably compute correlators along RG flows of quantum
field theories using the AdS/CFT correspondence. In particular, it
allows one to compute one-point functions in the presence of sources, to
check that they obey the requisite Ward identities throughout the
flow, and to compute power-law terms which were usually dropped in the
old prescription~\cite{GuKlPo98,FMMR98}.  In a theory that is
conformal also in the infrared, such as the one considered here,
surely a basic requirement of the formalism one wishes to apply is
that it is sufficiently restrictive to single out the correct
asymptotic power-law behavior automatically.  This is true for
holographic renormalization, as we show for example in
section~\ref{sec:2ptinert}.  As a consequence of being able to compute
one-point functions, one can also make sure one is using a
renormalization scheme that preserves supersymmetry, since $\langle
T_{ij}\rangle=0$ is then expected to hold in the background.  Finite
counterterms have to be added to ensure this, corresponding to a
selection of a renormalization scheme that preserves supersymmetry in
the dual field theory.

After this brief introduction to the formalism, we proceed to
summarize the new results in this paper.  We extend the analysis of
\cite{BiFrSk01,BiFrSk01a} 
to also include supermultiplets consisting of ``inert'' scalars $\Phi$
(meaning they have vanishing background --- as opposed to the
``active'' scalar $Q$ which carries the domain wall background). As it
turns out, the treatment of the inert scalars raises some new
conceptual questions, such as the proper choice of finite counterterms
and the distinction between inert scalars with conformal dimensions
$\Delta_+$ and $\Delta_-$, corresponding to the same mass in
supergravity, as we shall shortly discuss.
Through the coupling to the active scalar $Q$, the inert scalars
change mass along the flow to the fixed point. This coupling also
allows for (in fact, requires) a new type of finite counterterm, of
the form $Q^2 \Phi^2$.  
We compute the coefficients of these terms
for all inert scalars, as well as the
coefficient of the finite counterterm $Q^4$ (first 
displayed in \cite{BiFrSk01}). 
The terms of type $Q^2 \Phi^2$
can be seen as natural generalizations of
the finite $Q^4$ term,
but the coefficients of the $Q^2 \Phi^2$ terms
cannot be determined by evaluation on the background, since
$\Phi$ itself vanishes on the background. In section
\ref{sec:susyWI} we will see how to compute 
those coefficients, using a
supersymmetry Ward identity for the two-point functions of
superpartner inert scalars; the result is listed in
\Ref{tableb}.
By adding all counterterms and taking the cutoff
$\epsilon$ back to zero, we compute the renormalized action.
As usual, all correlators are computed by 
functional differentiation of this renormalized
action with respect to the boundary sources,
and then setting the sources to zero.

The next result is more conceptual. It was noticed in
\cite{BerSam01} that the effective potential $\CV$ appearing in the
fluctuation equations of the inert scalars can be expressed in terms
of a simple prepotential in the sense of supersymmetric quantum
mechanics (susy-QM).\footnote{In this paper, 
the notion of {\em prepotential}
always refers to this supersymmetric quantum mechanics function $\CU$
encoding the potential of a fluctuation equation as $\CV=\CU'+\CU^2$,
as opposed to the {\em superpotential} $W$ which describes the
background potential $V$ of the active scalar as $V=\ft12
(W')^2-2W^2$.} This property 
is very useful, since it guarantees the
absence of tachyonic fluctuations.
Similarly, such prepotentials were found in the known five-dimensional
examples \cite{DWoFre00}; the general existence of
susy-QM prepotentials
seemed in need of
further explanation. In section~\ref{sec:pre} we show that the
existence of susy-QM prepotentials follows from the preserved $N=1$
supersymmetry of the background flow, and that the prepotentials may
be directly extracted from the fermionic mass term of the underlying
gauged supergravity. This argument extends readily to higher
dimensions.  This is one aspect in which we see that the toy model
does provide useful information in higher dimensions.

We then proceed to determine the supersymmetry Ward identity relating
two-point functions of a pair of superpartner scalars. This is useful
as it provides a way to distinguish between different conformal
dimensions associated to the same supergravity mass. It is well known
that scalars of mass in a certain range (here $-1 < m^2 \le 0 $) can
correspond to two different solutions $\Delta_{\pm}$ for the conformal
dimension of the dual operator. For the active scalar, the choice of
 $\Delta_+$ or $\Delta_-$ is the difference between whether the
background describes an operator flow or a vev flow~\cite{KleWit99},
but this is not so for the inert scalars. Our scalars are
precisely in this range, and unlike in previously studied cases, there
are now two representation sectors with the same quantum numbers, so
group theory is not sufficient to make the distinction.
Using the fact that the correlator asymptotics
depends directly 
on $\Delta$ and not just on the supergravity mass, 
and following the correlators from one fixed point to the other,
we are able
to distinguish between $\Delta_+$ and $\Delta_-$.

We then derive the fluctuation equations for the inert scalars, active
scalar and metric around the domain wall solution.  
In all previously studied cases, those fluctuation
equations were hypergeometric. Instead, we find that around our domain
wall solution all fluctuation equations, for inert and active scalars
as well as for the metric and vector fields, reduce to a slightly more
complicated equation, the {\it biconfluent Heun equation}, which
descends by confluence
from a Fuchsian equation with four regular singularities. We
devise some methods to solve this equation; the mathematics is
relegated to Appendix~\ref{AHeun}
where in particular we point out a simple and efficient way to compute
the sought-after coefficient numerically. This allows us to achieve
our main goal: the computation of two-point correlation functions
throughout the renormalization group flow.

The final part of the paper concerns the computation of a $C$~function,
i.e.\ a function that is monotonic as a function of RG scale along the
flow and interpolates between the central charges at the conformal
fixed points. The general
existence of such a function can be very useful to
map out the space of field theories, for instance to find well-defined
universality classes. In two dimensions, Zamolodchikov has given a
general construction in terms of the two-point correlators of the
stress-energy tensor \cite{Zamo86}. Since Zamolodchikov's proof of
monotonicity relies heavily on the lack of distinct tensor structures
for stress-energy 2-point functions in two dimensions, it has no
straightforward generalization to higher dimensions, but there have
been several proposals for defining monotonic $C$~functions by
holography. In particular, in \cite{GPPZ98,FGPW99} positive-energy
conditions in the bulk were used to produce a monotonic boundary
function in terms of the superpotential of the flow. It is now
interesting that in our two-dimensional example, we have both these
objects at our disposal: Zamolodchikov's $C$~function in terms of the
holographic correlators, as well as the holographic proposal of
\cite{GPPZ98,FGPW99}. 

The paper is organized as follows.  In section~2 we review the exact
domain wall solution of \cite{BerSam01} that describes the flow.  In
section~3, we display the field equations and solve them
perturbatively, i.e.\ close to the AdS boundary.  In section~4, this
solution is used to compute counterterms to form the renormalized
action, which is then functionally differentiated to give the
one-point functions. Solving the bulk
fluctuation equations, this is
already sufficient information to determine the two-point functions for
inert scalars in section~5. In this section we also address the
conceptual issues of prepotentials and distinction between $\Delta_+$
and $\Delta_-$. For the sector of active scalar and stress-energy
tensor fluctuations, one needs to do some more work, since their
mutual coupling requires one to find gauge invariant quantities to
work with. This is done in
section~6 and we compute
their linearized fluctuation equations around the domain wall. Then,
in section~7, we use the stress-energy correlators to study the
Zamolodchikov $C$~function and contrast it with other proposals for a
$C$~function.

Since this paper is relatively long, let us give some guidelines on
how the reader can get the most out of it in the shortest amount of
time, depending on his or her preferences.

\begin{itemize}
\item
{\it Learn the formalism of holographic renormalization}: this paper
provides an example that is technically less demanding than in the
defining papers \cite{dHSoSk00,BiFrSk01,BiFrSk01a}, where emphasis was
on the 5d/4d case, yet is still quite different 
from the basic example of a free massive scalar in the review
 \cite{Sken02}. The reader with
this interest would be well-advised to concentrate on sections
\ref{sec:eom} and \ref{sec:ctr}; the philosophy is explained in the
latter, with the core explanation in section~\ref{sec:Sren}.  Then it
is straightforward to derive inert correlators as in
\ref{sec:2ptinert}; the active/metric sector also requires decoupling
their fluctuation equations, as in section \ref{sec:active}.

\item
See {\it new conceptual issues that are also relevant in other
dimensions}, like the inclusion of inert scalars, the discussion of
prepotentials, and the distinction between $\Delta_+$ and $\Delta_-$
using correlators.  These issues are dealt with in
section~\ref{sec:twopoint}.  The notation is fairly standard in the
literature so skipping previous details should not encumber the expert
reader much.

\item
Readers interested mainly in {\it deformations of CFT, not
supergravity details} can enjoy our main results, the deformed 2-point
functions, in section~\ref{sec:2ptinert} and at the end of
section~\ref{sec:2ptactive}.  The notation is mostly self-explanatory,
but it might be wise to simultaneously consult appendix B where
$\Psi_{\alpha}(p)$ and the relevant special functions are
explained. The $C$ function is presented in section~\ref{sec:C}.

\end{itemize}

\section{An exact holographic conformal-to-conformal flow} 
\label{sec:summary}

In this section, we briefly review the analytic domain wall solution
of~\cite{BerSam01} which interpolates between two AdS vacua. It was
constructed as a solution in the three-dimensional $N\equ8$ gauged
supergravity with local $SO(4)\cro SO(4)$ symmetry~\cite{NicSam01b},
describing the $AdS_3\times S^3 \times S^3 \times S^1$ near-horizon
geometry of the double D1-D5 system~\cite{BoPeSk98,dBPaSk99} with
equal D5-brane charges. The matter sector of this theory consists of
$n$ multiplets each containing $8$ scalars and $8$ fermions, whereas
graviton, gravitini, and the $12$ vector fields are non-propagating in
three dimensions. The $8n$ scalars parametrize the coset manifold
$SO(8,n)/(SO(8)\cro SO(n))$.

The supergravity Lagrangian is given by~\cite{NicSam01b}
\ba
\CL&=& \ft14 \,\sqrt{G}R
+ \CL_{\rm CS} + 
\ft14\,\sqrt{G} G^{\mu\nu}\,\CP_\mu^{Ir} \CP_\nu^{\,Ir} 
+  \sqrt{G}\,V + \CL_{\rm F}\;,
\la{L}
\ea
where $\CL_{\rm CS}$ is the Chern-Simons term for the vector fields,
the third term is the kinetic term for the scalars, given 
explicitly in eq.\ (\ref{currpq}) below, 
$V$ denotes the scalar potential and finally
$\CL_F$ contains the fermionic terms, given in~\cite{BerSam01}.
We use indices $I, J, \dots$ and indices $r, s,
\dots$ to label the vector representations of $SO(8)$ and $SO(n)$,
respectively. The 12 vector fields transform in the adjoint
representation of the gauge group
\be
SO(4)^+\times SO(4)^- ~\subset~ SO(8) ~\subset~ 
SO(8)\times SO(n) ~\subset~ SO(8,n)\;,
\la{gaugeG}
\ee
where we use superscripts $\pm$ to distinguish the two
three-spheres. In addition to the local gauge symmetry, the theory is
invariant under the rigid action of $SO(n)$. Assuming $n\ge4$ matter
multiplets, we break the latter down to $SO(4)\times SO(n\mis4)$ and
consider the following subgroup
\ba
G_{\rm inv} &\equiv& SO(4)_{\rm inv}\times SO(n\mis4) 
\non 
&\subset&
\left(SO(4)^+ \times SO(4)^- \right) 
\times \left( SO(4)^{\vphantom +} \times SO(n\mis4) \right) 
\;,
\la{Ginv}
\ea
of the global invariance group of the potential $V$. The $SO(4)_{\rm
inv}$ factor in $G_{\rm inv}$ is embedded as the diagonal of the three
$SO(4)$ factors on the right hand side. 
Evaluation of the scalar potential $V$ on the two-dimensional space of
singlets under $G_{\rm inv}$ leads to
the potential~\cite{BerSam01}
\ba
V&=& -g^2\, \Big( 16+24\,(Z_1^2\pls Z_2^2) +8 \,(Z_1^2\pls Z_2^2)^2 - 
8\,(Z_1^6\pls Z_2^6) - 4\, (Z_1^4\pls Z_2^4)^2 \Big)
\;.
\la{Vxy}
\ea
where $g$ is the remaining gauge coupling constant.

\begin{figure}[htbp]
  \begin{center}
\epsfxsize=65mm
\epsfysize=65mm
\epsfbox{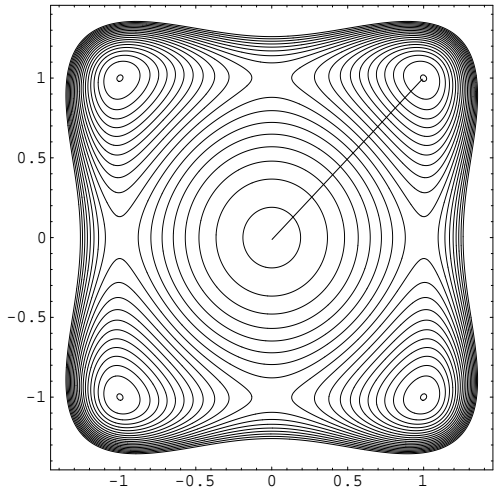}
  \caption{\small 
Contour plot of the scalar potential $V(Z_1,Z_2)$ \Ref{Vxy} and the
flow trajectory. }
 \label{Vpic}
  \end{center}
\end{figure}

The form of this potential is depicted in figure~\ref{Vpic}. It
exhibits two inequivalent extremal points apart from the local maximum
at the origin. The saddle point at $(Z_1,Z_2)=(1,0)$ corresponds to a
nonsupersymmetric but stable vacuum as has been verified by explicit
computation of the scalar fluctuations around this
point~\cite{BerSam01}. We will concentrate on the extremum located at
$(Z_1,Z_2)=(1,1)$, which preserves $N=(1,1)$ supersymmetry.

The ratio of the central charges of the dual conformal field theory at
this extremum and that of the CFT at the origin is given by
\cite{BroHen86,HenSke98}
\be
\frac{c_{\rm IR}}{c_{\rm UV}} 
~=~ \sqrt{\frac{V_{\rm UV}}{V_{\rm IR}} } ~=~ \frac12
\;,
\la{ccVV}
\ee
supporting the conjecture that this point corresponds to a mass
deformation of the UV conformal field theory; half the fields are
integrated out to form the IR theory.  The supergravity spectrum
around this point is organized in $N=(1,1)$ supermultiplets as
summarized in table~\ref{specL0}, where $h$ and $\bar{h}$
denote the conformal dimensions associated with the supergravity
masses, so that $\Delta= h+\bar{h}$.  Note that the
multiplet in the ${\bf (1,1)}$ for instance contains two scalars and
spin-$\ft12$ fields, whereas the multiplet in the ${\bf (1,3)}$
combines a spin-$\ft12$ field with massive selfdual vectors and a
massive gravitino.

\begin{table}[htb]
\centering
\begin{tabular}{|c||c|c|} \hline
$SO(4)_{\rm inv}$  &
\begin{tabular}{c} $N=(1,1)$ multiplets \\
$h \;\;\, \times \;\;\, \bar{h}$
\end{tabular}  & 
\begin{tabular}{c} field content \\
$ (h,\bar{h})$ 
\end{tabular} \\
\hline\hline
{\bf (1,1)}  & $(\ft54|\ft74)\times(\ft54|\ft74)$ 
& $(\ft54,\ft54)$, $(\ft74,\ft74)$, $(\ft54,\ft74)$, $(\ft74,\ft54)$\\
\hline 
{\bf (3,3)} & $(\ft14|\ft34)\times(\ft14|\ft34)$ 
& $(\ft14,\ft14)$, $(\ft34,\ft34)$, $(\ft14,\ft34)$, $(\ft34,\ft14)$\\
\hline 
{\bf (1,3)} & $(\ft14|\ft34)\times (\ft54|\ft74)$ 
& $(\ft14,\ft74)$, $(\ft14,\ft54)$, $(\ft34,\ft74)$, $(\ft34,\ft54)$\\
\hline 
{\bf (3,1)} & $(\ft54|\ft74)\times (\ft14|\ft34)$ 
& $(\ft74,\ft14)$, $(\ft54,\ft14)$, $(\ft74,\ft34)$, $(\ft54,\ft34)$\\
\hline 
{\bf (2,2)} & $(\ft12|1)\times(\ft12|1)$ 
& $(\ft12,\ft12)$, $(1,1)$, $(\ft12,1)$, $(1,\ft12)$\\
\hline 
\end{tabular}
\caption{\small Supergravity spectrum 
around the supersymmetric vacuum $(Z_1,Z_2)=(1,1)$.}
\label{specL0}
\end{table}

An analytic domain wall solution interpolating between the origin and
the supersymmetric extremum was constructed
in~\cite{BerSam01} using the ansatz
\be
Z_1 ~=~ Z_2 ~=~ \ft1{\sqrt{2}}\,\sinh
\left(\ft1{\sqrt{2}}Q\right) \;,
\la{Q}
\ee
where $Q$ denotes the active scalar field parametrizing the diagonal
in figure~\ref{Vpic}. The remaining scalar fields are collectively
referred to as {\em inert} scalars. Around the origin $Z_1=Z_2=0$, the
mass of the active scalar field is $m^2L_0^2=-\ft34$, i.e.\ it is dual
to a relevant operator of conformal dimension $\Delta=\ft32$ which
drives the flow away from the UV conformal field theory. In the
truncation \Ref{Q}, the bosonic part of the Lagrangian~\Ref{L} reduces
to
\ba
\CL &=& \ft14\, \sqrt{G} R 
+ \ft12\,\sqrt{G} G^{\mu\nu}\, \partial_{\mu}Q\,\partial_{\nu}Q
+V_Q \;,
\la{Lbos}
\ea
with a scalar potential derived from a superpotential $W$ as
\ba
V_Q  &=& \frac12 \,(\dd_Q W)^2 - 2 W^2\;,
\quad
W\equiv -\frac{g}{8}\, \left(13 + 20 \cosh (\sqrt{2}Q) - 
\cosh (\sqrt{8}Q)\right) \;.
\la{VQ}
\ea
With the standard domain wall ansatz
\be
\la{dwmetric}
ds^2~=~e^{2A(r)}\,\eta_{ij}\,dx^i dx^j + dr^2 \;,
\ee
the field equations may be reduced to first-order form: 
\be
\dd_r Q ~=~ \dd_Q W \;,
\qquad
\dd_r A ~=~-2\,W \;.
\la{DW}
\ee
They may analytically be solved by
\be
\frac{(5-y)(y+1)^2}{16\,(y-1)^3} ~=~  e^{24gr} \,,
\quad
e^{6A(r)} ~=~ \frac{(5-y)^{4}}{128\,(y+1)(y-1)^6} \,,
\quad
y=\cosh (\sqrt{2} Q) \;.
\la{kink}
\ee
This solution interpolates between the origin and the supersymmetric
extremum, preserving $N=(1,1)$ supersymmetry throughout the flow. From
now on, we will fix the gauge coupling constant to the numerical value
$g=1/8$, thereby setting $L_0$, the AdS length at the origin of the
scalar potential, to unity.

Let us emphasize that for the computation of correlation functions,
it is indispensable to have an {\em exact}, meaning analytic, domain wall
solution \Ref{kink}, rather than just finding the first-order
equations and solving them numerically, which may be sufficient for
other purposes. The point is that it is the fluctuations of
supergravity fields {\it around the domain wall solution} that contain
information on correlators in the boundary field theory. In
particular, these second-order fluctuation equation are supplied with
boundary conditions that specify the value of a supergravity field at
the AdS boundary and demand a given (regular) behavior in the deep
interior of the bulk space. The latter condition cannot be fixed
perturbatively from the boundary, but is fixed only by solving the
fluctuation equations. One might say this requires the
``nonperturbative'' bulk information that encodes the two-point
function.\footnote{Of course, ``perturbative'' here refers to the
radial expansion, which is a derivative (low-energy) expansion, rather
than the coupling constant expansion in the field theory.}  The
fluctuation equations we obtain are highly singular at infinity, and
it is not clear that one could have obtained our correlators even
numerically if our domain wall solution would have amounted to just a
numerical solution of the first-order equations.

\section{Field equations and near boundary analysis}
\label{sec:eom}

Thus, we embark on the way to computing correlators using the
domain wall solution given above. As a first step we will derive the
supergravity field equations and determine the coefficients in the
near-boundary expansion of the supergravity fields. The truncated
supergravity Lagrangian is given in \Ref{Lbos}. In principle, it
contains the complete information to compute correlation functions of
the stress-energy tensor along the flow. Since we will be interested
in computing correlation functions of the operators associated with
the inert scalars as well, we extend \Ref{Lbos} by expanding the
original Lagrangian \Ref{L} to second order in the inert scalars. This
is sufficient for all further computations in this paper, as we will
only treat correlation functions with at most two insertions of inert
scalar operators. We denote the inert scalars collectively by
\ba
\Phi^{\bf i} &=& \left \{ \Phi^{\bf 1},\Phi^{{\bf 9}+},
\Phi^{{\bf 9}-},\Phi^{{\bf 4}+},\Phi^{{\bf 4}-} \right\} \;,
\ea
where the index ${\bf i}$ denotes the dimension of the representations
under $SO(4)_{\rm inv}$ while the superscripts $\pm$ label the
two-fold degeneracies, 
cf.\ the spectrum in table~\ref{specL0}. Recall that
the scalar fields in the ${\bf (3,1)}+{\bf (1,3)}$ appear in a
multiplet together with the vector fields to which they are related by
gauge symmetry. Hence, these scalars require a separate analysis and
will not be treated in this paper.

The scalar fields parametrize an $SO(8,n)$ matrix according
to\footnote{To avoid confusion, we will always explicitly write out
the sums over the different representations labelled by ${\bf i}$.}
\ba
\Mat&=&\Mat_Q\,\Mat_\Phi\;,\qquad
\mbox{with}\quad \Mat_\Phi = 
\exp \sum_{\bf i} \Phi^{\bf i}\, Y^{\bf i}  \;,
\la{infsc}
\ea
where $\Mat_Q$ carries the entire dependence on the active scalar $Q$,
and $Y^{\bf i}$ denote the noncompact generators of $SO(8,n)$
associated with the representations of $\Phi^{\bf i}$, see
\cite{NicSam01b,BerSam01} for details. With this ansatz, the current
in the kinetic term of \Ref{L} becomes
\ba
\CP_\mu^{Ir}\,Y^{Ir} &=& 
\Mat^{-1}\,D_\mu\Mat ~=~
\Mat^{-1}_\Phi\,D_\mu\Mat^\vl_\Phi + 
\Mat^{-1}_\Phi
\left(\Mat^{-1}_Q D_\mu\Mat^\vl_Q\right)\,\Mat^\vl_\Phi \; ,
\la{currpq}
\ea
where the $I$, $r$ indices belong to the vector representations of
$SO(8)$ and $SO(n)$, respectively, and the relation to the $\Si$
representation sector index is as described in
section~\ref{sec:summary}.

The near-boundary analysis is most conveniently performed in 
Fefferman-Graham coordinates $(x^i, \rho=e^{-2r})$, taking the
metric to be
\be
ds^2 = {1 \over 4\rho^2}\, d\rho^2 
+ {1 \over \rho}\, g_{ij} dx^i dx^j
\qquad i=1, \ldots , d \;,
\label{eq:metric}
\ee
where for the moment we let $d$, the dimension of the boundary, be
arbitrary for the sake of easy comparison with standard
literature. The action 
on the asymptotically AdS space $M$ 
is then of the general form
\ba
S &=& \int_{M}
d^d x \, d\rho \, \sqrt{G} \,
\Big(\ft1{2\kappa}R + \ft12 G^{\mu\nu}
K(\Phi)\, \dd_\mu Q \dd_\nu Q 
+ \ft12 G^{\mu\nu}\sum_{\bf i} 
\partial_{\mu}\Phi^{\bf i}\partial_{\nu}\Phi^{\bf i} 
+V(Q,\Phi)
\Big) \non
&&{}
- \ft12 \int_{\partial M} d^d x \, \sqrt{\gamma} \,{\cal K} \; ,
\label{eq:action}
\ea
for general gravitational coupling constant $\kappa$.
We denote by $\CK$ the trace of the
extrinsic curvature tensor $\CK_{ij}$ on the hypersurface $\partial
M$, and $\CK_{ij}$ itself is  given by
\be
\CK_{ij} = {1\over \rho}\, g_{ij} - \partial_{\rho} g_{ij}
\;,
\label{eq:extrinsic}
\ee
in the metric (\ref{eq:metric}).  The induced metric on the
hypersurface $\partial M$ is denoted $\gamma_{ij}$.  The functions
$V(Q,\Phi)$ and $K(\Phi)$ in \Ref{eq:action} are obtained from
expanding (to second order in $\Phi$) the scalar potential and the
kinetic term, respectively, of~\Ref{L}.\footnote{It is due to our
parametrization \Ref{infsc}, \Ref{currpq} that the inert scalars
automatically arise with a canonical kinetic term, whereas
the kinetic term of the active scalar $Q$ depends
on the inert scalars. In
contrast, the commonly used parametrization 
(which corresponds to choosing $\Mat=\Mat_\Phi\Mat_Q$ in \Ref{infsc}) 
gives a canonical kinetic term for the active scalar and a
$Q$-dependent metric for the inert scalars. This requires additional
rescaling of the inert scalars  in order to diagonalize their equations
of motion, cf.~\cite{BWFP01}, which essentially amounts to reverting to
the parameterization \Ref{infsc}.}  Somewhat miraculously, the quadratic
parts of the two functions may simultaneously be diagonalized with
$Q$-independent eigenvectors in each of the two-fold degenerate
representation sector~ ${\bf 4}$ and~${\bf 9}$, respectively, and take
the form
\ba
V(Q,\Phi)=V_Q+\sum_{\bf i} V_{\bf i}(Q)\,\Phi^{\bf i}\Phi^{\bf i}
\;,\qquad
K(\Phi)= 1 + 
\frac12\,\sum_{\bf i} K_{\bf i}\,\Phi^{\bf i}\Phi^{\bf i} \;,
\la{KVi}
\ea
with
\ba
K_{\bf 1}=K_{{\bf 9}-}=K_{{\bf 4}-}=1\;,&& \quad
K_{{\bf 9}+}=K_{{\bf 4}+}= 0\;,
\la{Ki}
\ea
and the $V_{\bf i}(Q)$ given in \Ref{Vi0}, \Ref{Vi} below. The
equations of motion from varying this action are, for a general metric
$G_{\mu \nu}$\,\footnote{Our curvature conventions are
$R_{\mu\nu\sigma}{}^{\tau}=\dd_\mu\Gamma^\tau_{\nu\sigma} +
\Gamma^\tau_{\mu\lambda}\,\Gamma^{\lambda}_{\nu\sigma} -
(\mu\leftrightarrow\nu),\,R_{\mu\nu}=R_{\mu\tau\nu}{}^{\tau}$.}
\ba
R_{\mu\nu}[G]+{2 \over d-1}\,\Lambda G_{\mu\nu} &=& 
-\kappa\Big[ K\, \dd_\mu Q \dd_\nu Q 
+ \sum_{\bf i} 
\partial_{\mu}\Phi^{\bf i} \partial_{\nu}\Phi^{\bf i} +
{2G_{\mu\nu}\over d-1}\,(V-V_{(0)}) \Big] \;, \non
\square_G \,\Phi^{\bf i} &=&
{\partial \over \partial \Phi^{\bf i}} 
\left(
V + \ft12 K G^{\mu\nu} \dd_\mu Q \dd_\nu Q \right)\;,
\label{eq:generalEOM}
\\[1ex]
\Ksquare_G \, Q &=&
{\partial V \over \partial Q} \;, \nonumber
\ea
where $V_{(0)}= V(Q,\Phi)|_{Q=0,\Phi^{\bf i}=0}$, and $\Lambda=\kappa
V_{(0)} = -{d(d-1) \over 2}$\,, and $\Ksquare_G Q$ is 
the expression obtained by varying 
$\ft12 \sqrt{G}K G^{\mu\nu}\partial_{\mu}Q\partial_{\nu}Q$
and discarding a boundary term.
 With the metric~\Ref{eq:metric}, the
Einstein equations can be written in the following form,
\ba
\rho[2 g_{ij}'' -2(g'g^{-1}g')_{ij}+\Tr(g^{-1}g')g'_{ij}]
+R_{ij}[g]  
 -(d\mis2)\,g_{ij}'-\Tr(g^{-1}g')g_{ij} ~ = 
\hspace*{1.2cm}\non[.5ex]
-\kappa\Big[ K\, \dd_i Q \dd_j Q 
+ \sum_{\bf i} \partial_i\Phi^{\bf i} \partial_j\Phi^{\bf i} +
{2\over (d\mis1)}{g_{ij}\over \rho} \,  (V-V_{(0)}) \Big]  \;,
\nn
\ea
\vspace*{-.7cm}
\ba
\nabla\!_i \, \Tr (g^{-1} g') -  \nabla^j g_{ij}' &=& 
-2\kappa\, \Big[K\, Q' \dd_i Q 
+ \sum_{\bf i} \Phi^{\bf i}{\,}' \partial_i \Phi^{\bf i}\Big]  \;, 
\la{Einstein}
\\
\Tr(g^{-1} g'') - \ft12 \Tr (g^{-1} g' g^{-1} g') 
&=& -2\kappa\, \Big[K\, (Q')^2
+ \sum_{\bf i} (\Phi^{\bf i}{\,}')^2 + {V-V_{(0)} \over
2(d\mis1)\rho^2} \;  \Big] \;,
\nn
\ea
and we note that here $R_{ij}[g]$ means the Ricci tensor for the
$d$-dimensional metric $g_{ij}$ only, i.e.\ only ${R_{ikj}}^k$ and no
${R_{i\rho j}}^{\rho}$ piece.  The scalar equations take the form
\ba
4\rho^2 \Phi^{\bf i}{}\,'' + 2\rho\Phi^{\bf i}{}\,'
\left[\left(2\mis d\right)
+ \rho \, (\log g)'\right] 
+ \rho\, \square_g \Phi^{\bf i}{} ~=
\hspace*{6cm}&& \non
\left(2V_{\bf i} + \ft12\,K_{\bf i}
(4\rho^2 Q'Q' + \rho g^{ij}\dd_i Q \dd_j Q) \right) \Phi^{\bf i} 
\;,&&
\nn
\ea
\vspace*{-.9cm}
\ba
4\rho^2 (KQ''+K'Q') + 2\rho KQ'\left[\left(2\mis d\right)
+ \rho \, (\log g)'\right] 
+ \rho\, \Ksquare_{g}\, Q - \partial_Q V  &=& 0 \;.
\label{eq:scalard}
\ea
Primes here denote derivatives with respect to $\rho$, and $g=\det
g_{ij}$. From now on, we specialize to $d=2$, but at times we will
return to compare our results to those of higher dimensions, $d=4$ in
particular.  For reference, it will be useful to have the first
Einstein equation for $g_{ij}=f(\rho)\,\eta_{ij}$:
\be
2\rho^2 f'' - 2\rho f' + 2\kappa\, (V-V_{(0)}) \,f = 0 \;,
\label{eq:verify1}
\ee
and the scalar equation of motion for $x$-independent
$Q=Q(\rho)$ in this metric, with $\Phi^{\Si}=0$:
\be
4\rho^2\,\Big({f' \over f} Q' + Q''\,\Big) = {\partial V \over
\partial Q} \;.
\label{eq:verify2}
\ee

\subsection{Perturbative solution}

We will now solve the above equations of motion perturbatively for the
first few terms in the expansion in $\rho$. First, we note that the
kink solution~\Ref{kink} admits an expansion according to
\ba
Q_{\rm B} &=& 
%\rho^{1/4}\,\acq ~=~ 
%\rho^{1/4} \, \sum_n \acq_{(n)}\, \rho^{{n/2}} ~=~
\rho^{1/4}
\,\Big(1+ {1 \over 24} \sqrt{\rho} -{13 \over 640}\,  \rho + 
\ldots\,\Big) \;,
\non[.5ex]
G_{{\rm B}\,ij} &=&
%\rho^{-1} \, g_{ij} ~=~
%\rho^{-1} \,  \sum_n g_{(n)\,ij}\, \rho^{{n/2}} ~=~ 
\rho^{-1}\,\eta_{ij}\,  \Big(1-
\sqrt{\rho} + {7 \over 16}\,  \rho + \ldots \Big)  
\;,
\la{BGexpansion}
\ea
where ``B'' is for background.  Consistency of these expansions
can be conveniently checked using (\ref{eq:verify1}) and
(\ref{eq:verify2}).  The appearance of square roots of $\rho$ in these
expansions is generic in three bulk dimensions.  One could have chosen
to expand in a different variable to avoid the square roots, but we
find it more useful to stay notationally close to the
higher-dimensional literature.  In general, the metric and scalars
may also depend on $x$, and we expand the fields in $\rho$ as
\ba
G_{ij}(x,\rho) &=& \rho^{-1} g_{ij}(x,\rho) \label{expansions}\\
&=& 
\rho^{-1} \Big\{ 
g_{(0)\,ij}(x) + \sqrt{\rho} g_{(1)\,ij}(x) + \rho \left(
g_{(2)\,ij}(x) + h_{(2)\,ij}(x)\log \rho  \right) + \dots \Big\} 
\;,
\non[.5ex]
Q(x,\rho) &=& \rho^{1/4} q(x,\rho) ~=~ 
\rho^{1/4} \Big\{ 
q_{(0)}(x) + \sqrt{\rho} \left( q_{(1)}(x) + 
\tilde{q}_{(1)}(x)\log \rho 
\right) + \dots \Big\} \;,
\non[.5ex]
\Phi(x,\rho) &=& \rho^{1/4} \phi(x,\rho) ~=~ 
\rho^{1/4} \Big\{ 
\phi_{(0)}(x) + \sqrt{\rho} \left( \phi_{(1)}(x) + 
\tilde{\phi}_{(1)}(x)\log \rho 
\right) + \dots \Big\} \;.
\nn
\ea
The subscripts in parenthesis denote the order in $\sqrt{\rho}$, and
also the highest number of derivatives with respect to $x$
that will occur in these coefficients, hence this is a derivative
(low-energy) expansion. The coefficients of the leading terms
$g_{(0)\,ij}$, $q_{(0)}$, $\phi_{(0)}$ are interpreted as source terms
for the dual operators. The field equations then determine the next
few coefficients as algebraic functions of these boundary data. At a
certain order --- here, $g_{(2)}$ for the metric and $\phi_{(1)}$
for the scalars --- the desired coefficient cancels out of the
perturbative equations of motion, and remains undetermined; it is
related to the one-point function
of the dual operator. At this order, 
the ansatz is generalized to include a logarithmic term as
we have done in \Ref{expansions}, 
and the coefficient of this logarithmic term is determined instead
of the one that cancelled out. 
In general,
for example in the $d=4$ Coulomb branch (CB) flow (see e.g.\
\cite{BiFrSk01}), one may also need higher powers of logarithms
in the ansatz, but we have checked that we do not.

The expansion of the scalar potential $V_Q$ gives
\ba
V_Q &=& -\frac12 -\frac38\,q_{(0)}\,
\sqrt{\rho} -\frac18\,(q_{(0)}^4 + 6 q_{(0)}
q_{(1)})\,\rho  - \frac34\,q_{(0)}\tilde{q}_{(1)} \,\rho\log\rho + \dots
\;,
\ea
while for the inert scalar potentials $V_{\bf i}$ we find
\ba
V_{\bf i} &=& -\frac38 
- \frac14\,(1+\ft14 K_{\bf i})\, q_{(0)}^2\,\sqrt{\rho} + \dots
\;.
\ea
To this order, the potentials $V_{\bf i}$ come in just two different
forms, depending on whether the value of $K_{\bf i}$ \Ref{Ki} is 0 or
1. It is only at higher order that the potentials $V_{\bf i}$ begin
to differ between the various inert scalars.
We can now expand the full potential as
\[
V(Q,\Phi) = V_{(0)}+\sqrt{\rho}\;  V_{(1)}(x)
+\sqrt{\rho} \log \rho \;  \tilde{V}_{(1)}(x) +
\rho\;  V_{(2)}(x)+\rho \log \rho \;  \tilde{V}_{(2)}(x) + \ldots
\]
and find for the coefficients
\ba
V_{(0)}&=&-\frac12 \; ,
\qquad
 V_{(1)} ~=~ -\frac38\, \Big(q_{(0)}^2 + 
\sum_{\bf i} (\phi^{\bf i}_{(0)})^2\Big)  \; ,
\qquad
 \tilde{V}_{(1)} ~=~ 0  \; ,
\non
V_{(2)} &=& -\frac18\, q_{(0)}^4 - 
   \frac14 \sum_{{\bf i}}  
(1+\ft14 K_{\bf i})\,q_{(0)}^2 (\phi^{\bf i}_{(0)})^2
-\frac34\, \Big( q_{(0)} q_{(1)} + \sum_{\bf i} \phi^{\bf
i}_{(0)}\phi^{\bf i}_{(1)}  \Big) \; ,
\non
 \tilde{V}_{(2)} &=&
-\frac34\, \Big(q_{(0)}  \tilde{q}_{(1)} 
+ \sum_{\bf i}\phi^{\bf i}_{(0)} \tilde{\phi}^{\bf i}_{(1)}\Big) 
\;.
\la{V1}
\ea
To lowest order, only the potential $V_Q$ for the active scalar
contributes. The fact that $\tilde{V}_{(1)}$ vanishes is required for
consistency; there is a $1/\rho$ in front of $V$ in the equation of
motion, giving a total of $\rho^{-1/2}\log \rho$ for the
$\tilde{V}_{(1)}$ term, but there is no term of that order to match it
on the left-hand side of Einstein's equation.

\subsection{Metric coefficients}
\label{sec:metriccoeff}

Solving the $(ij)$ component of the Einstein field equations
\Ref{Einstein} for each coefficient in (\ref{expansions}) leads to
\footnote{Here and in the following, we use the matrix notation
$\tr A \equiv \tr (g_{(0)}^{-1} A)$ and $A B \equiv A g_{(0)}^{-1} B$,
i.e.\ indices are raised and lowered with the boundary metric
$g_{(0)ij}$.}
\ba
g_{(1)ij} &=& \ft43 \kappa  V_{(1)}\, g_{(0)ij} \;,
\qquad 
\Tr g_{(2)} ~=~
\ft12 R[g_{(0)}] + 2 \kappa V_{(2)} + \ft83 \kappa^2 V_{(1)}^2
\;,
\non
h_{(2)ij} &=& \kappa \tilde{V}_{(2)}\, g_{(0)ij}  \;, 
\la{coeffmetric}
\ea
at orders $\rho^{-1/2}$, $\rho^{0}$, and $\log \rho$, respectively.
When one arrives at order $\rho^0$ of the Einstein equation, which
would determine the coefficient $g_{(2)ij}$, this coefficient only
appears traced on the left-hand side. Thus, the non-trace part of
$g_{(2)ij}$ remains undetermined in perturbation theory, which is
expected as remarked above.  Note that the last equation in
\Ref{coeffmetric} is understood to hold only up to terms in order
$\phi_{(0)}^4$, since our starting action \Ref{eq:action} was valid up
to this order, cf.~the discussion in the beginning of this section. It
is further worth pointing out that, in contrast to the known $d=4$
examples (e.g.\ the GPPZ flow), the $h_{(2)}$ coefficient in
\Ref{coeffmetric} is excited only by logarithmic terms in the scalars.
%\hmm{General statement about vanishing of logs in 2d?}

{}From the $(i\rho)$ component of Einstein's equations, one further
derives 
\ba
\nabla^j\,g_{(2)ij} 
&=& \nabla\!_i\,\Tr g_{(2)}  
-\frac38\,\nabla\!_i\,\Tr g_{(1)}^2 
-\frac14\,g_{(1)ij}\,\nabla^j\,\Tr g_{(1)}
\non
&&{}
+\frac12\,\nabla^j\,(g^2_{(1)ij})
+ \frac{\kappa}{2} 
\left( q_{(0)} \nabla\!_i\,q_{(1)}  +3\,q_{(1)} 
\nabla\!_i\,q_{(0)}  \right)
\non[.5ex]
&&{}
+\frac{\kappa}{2} \sum_{\bf i} \left(
\phi^{\bf i}_{(0)}\nabla\!_i\,\phi^{\bf i}_{(1)} + 
3\phi^{\bf i}_{(1)}\nabla\!_i\,\phi^{\bf i}_{(0)} +
\ft12 K_{{\bf i}}(\phi^{\bf i}_{(0)})^2 q_{(0)} \nabla\!_i\,q_{(0)} \right)
\;.
\la{trg2}
\ea
This expression will be used in section (\ref{sec:stressenergy})
to verify one of the Ward identities.

\subsection{Scalar coefficients}
\label{sec:scalarcoeff}

The equations of motion for the inert scalar fields (\ref{eq:scalard})
in $d=2$ read
\ba
2\rho^2  \Phi^{\bf i}{}\,'' + \rho^2\,(\log g)'\Phi^{\bf i}{}\,'
+ \ft12 \rho\, \square_g \Phi^{\bf i} &=&  
\left( V_{\bf i} + K_{\bf i}
(\rho^2 Q'Q' + \ft14\rho g^{ij}\dd_i Q \dd_j Q) \right) \Phi^{\bf i}
\;.
\la{eominert}
\ea
Expanding the right-hand side
of \Ref{eominert} in $\rho$, we find
\ba
-\frac34\,\rho^{1/4} \left( 
\phi^{\bf i}_{(0)} 
+ (\phi^{\bf i}_{(1)} + \ft23 \phi^{\bf i}_{(0)} q_{(0)}^2 ) 
\sqrt{\rho} 
+ \tilde{\phi}^{\bf i}_{(1)}  \sqrt{\rho}\log\rho
+ \dots \right) \;.
\ea
Remarkably, $K_{\bf i}$ drops out of the expression, i.e.\ to this order
in $\rho$, the effective potentials of all the inert scalar fields
coincide. 
Now expanding also the 
left-hand side of \Ref{eominert}, we find
that equating the most divergent terms 
(orders $\rho^{-1}$ and $\rho^{-1/2}\log \rho$) on the two sides
just gives consistency conditions. 
The next-to-leading-order
nonlogarithmic terms $\rho^{-1/2}$ have the same
dependence on $\phi_{(1)}$ on both sides, 
so it cancels out. With the logarithmic term in the ansatz
\Ref{expansions}, the coefficient $\tilde{\phi}_{(1)}$ also
appears at order $\rho^{-1/2}$, but it does not cancel and
is determined as promised:
\ba
\tilde{\phi}^{\bf i}_{(1)} 
&=& \frac13  \left( \Tr g_{(1)} + 2 q_{(0)}^2 
\right) \phi^{\bf i}_{(0)} ~=~ \CO(\phi^3) \;,
\ea
which upon using \Ref{coeffmetric} is of only cubic order in the inert
scalars, i.e.\ it vanishes to the order of validity of our computation
for all the inert scalars. As anticipated above, the coefficient
$\phi^{\bf i}_{(1)}$ remains undetermined by the field
equations. For the active scalar $Q$, one derives
from \Ref{eq:scalard} the analogous equation
\ba
\tilde{q}_{(1)} 
&=& -\frac18 \,  \Big( \Tr g_{(1)} + 2 q_{(0)}^2 + 2
\sum_{\bf i} (\phi^{\bf i}_{(0)})^2 
\Big) \, q_{(0)} ~=~ \CO(\phi^4) \;,
\ea
while $q_{(1)}$ remains undetermined as expected.  We are now in a
position to use these results to compute divergences of
the on-shell action and, from there, counterterms and one-point
functions.  From now on, we set $\kappa= 2$ for the gravitational
coupling constant.

\section{Counterterms and one-point functions}
\label{sec:ctr}

\subsection{Counterterms}
\label{sec:Sren}
As is well known, the on-shell action of gravity and scalars is a
priori divergent even on a background of pure AdS.  It contains
divergences coming from the $\rho\rightarrow 0$ (boundary) side, and
the standard regularization is to consider the action on a surface
away from the boundary $\rho=\epsilon$ for some small coordinate
distance $\epsilon$.  Then, divergences appear as poles 
and logarithms
in $\epsilon$
that can be subtracted, which corresponds to some subtraction
scheme in the boundary field theory. Finally, one can let $\epsilon
\rightarrow 0$ to obtain the renormalized action.  In addition to this
standard procedure, we also want to ensure the on-shell action
vanishes when evaluated on the background; this is necessary (but as
we will see in section \ref{sec:twopoint}, not sufficient in our case)
to make sure the renormalization scheme is supersymmetric
\cite{BiFrSk01,BiFrSk01a}. 
Since this formalism is not yet widely familiar, we will make an
effort to explain all steps carefully. 
(In this section, for transparency of the intermediate expressions, we
suppress the indices $\Si$ labeling the representation sectors of
inert scalars, and restore them at the end.) 
To start from the beginning,
the first step of holographic renormalization is one of convenience:
to eliminate the
Einstein-Hilbert term in the bulk action
against all scalar kinetic terms on-shell. This
is an amusing exercise that can be performed in any dimension, simply
by tracing the Einstein equation in \Ref{eq:generalEOM} and
substituting the Ricci scalar so obtained into the action
\Ref{eq:action}.  Only the potential $V$ then remains in the bulk
action, with the coefficient changed from $+1$ to $-2$, and there is a
factor of 1/2 from $\sqrt{G}=\sqrt{g}/2\rho^2$.  
Explicitly,
\ba
S_{\rm reg} &=& \int_{\rho \geq \epsilon}
d^d x \, d\rho \, \sqrt{G} \,
\Big(\ft1{2\kappa}R + \ft12 G^{\mu\nu}
K(\Phi)\, \dd_\mu Q \dd_\nu Q 
 \non [-1ex]
&& \qquad \qquad 
+ \ft12 G^{\mu\nu}\sum_{\bf i} 
\partial_{\mu}\Phi^{\bf i}\partial_{\nu}\Phi^{\bf i} 
+V(Q,\Phi)
\Big)- \ft12 \int_{\rho =\epsilon} 
d^d x \, \sqrt{\gamma} \,{\cal K} \; , 
\label{eq:Sreg} \\ [1ex]
&=& -\int  \int^{\rho_{\rm cr}}_{\epsilon}d^2 x\, 
 d\rho \; {\sqrt{g} \over \rho^2}\; V 
-\ft 12 \int_{\rho=\epsilon} 
d^2 x {\sqrt{g} \over \epsilon} \; \CK \;,
\label{eq:bulkreg}
\ea
where in the extrinsic curvature term we replaced $\sqrt{\gamma}=
\sqrt{g}/\epsilon$. Although expressions \Ref{eq:Sreg} and
\Ref{eq:bulkreg} are on-shell
equivalent, the action \Ref{eq:bulkreg} is only convenient
for determining counterterms and later we return to
using \Ref{eq:Sreg}. 
Here, the upper limit of integration $\rho_{\rm cr}(x)$ can be defined 
in general as giving the surface $\rho=\rho_{\rm cr}(x)$
of vanishing area, as in \cite{SkeSol99}, but for
a stationary metric as we consider here, all one needs is that the
bulk lapse and shift functions vanish at $\rho=\rho_{\rm cr}$.  In
the coordinates (\ref{eq:metric}), this simply means $\rho_{\rm
cr}=\infty$.

One then performs an
expansion in $\epsilon$ of \Ref{eq:bulkreg}.
Gravity without scalars in three bulk dimensions is expected to give a
constant volume divergence plus a finite term involving $\trgt$.
However, when we couple scalars we get all kinds of terms, including
$(\log \rho)^2$ due to $\Phi^2$, as we shall see.  We will use the
expansion of $\sqrt{\det g}$ to finite order:
\be
\sqrt{g} = \sqrt{g_{(0)}}\,
\bigg\{1+\sqrt{\epsilon}(  \ft12 \Tr g_{(1)}) + 
\epsilon (  \ft12 \Tr g_{(2)}) 
+\epsilon \log \epsilon\, (\ft12 \Tr h_{(2)})  
+ \CO(\epsilon^{3/2}) \bigg\}  \; .
\label{gexp}
\ee
Notice that the combination $\Tr g_{(1)}^2-\ft12(\Tr g_{(1)})^2$
vanishes for our solution, since $g_{(1)ij}$ is simply proportional to
$g_{(0)ij}$.  Substituting in all expansions, and staying with a
general potential, we arrive at
\ba
S_{\rm reg} &=& \int d^2 x \, \sqrt{g_{(0)}} \:\Big\{
\epsilon^{-1} (-V_{(0)} -1) 
+ \epsilon^{-1/2} (- \trgo V_{(0)} -2 V_{(1)} -\ft14 \trgo) 
\non
&& + \log \epsilon (
\ft12 \trgt
V_{(0)} + \ft12 \trgo V_{(1)} + V_{(2)} -\ft{d-2}4 \trh )\non [1mm] 
&& +(\log \epsilon)^2  \ft14(\tilde{V}_{(2)} 
+ \ft12 \trh  \tilde{V}_{(0)})  
+ \epsilon^0 (
 -\ft{d-2}4 \trgt  ) + \CO(\epsilon^{1/2}) \Big\}
\;.
\label{eq:withpot}
\ea
Of course we consider $d=2$, but we found it useful to keep 
coefficients of the form $(d-2)$ to display
some cancellations.  In particular, we see that the finite $\trgt$
contribution from the extrinsic curvature is cancelled.  Substituting
our perturbative solution for the metric (eq.\ \Ref{coeffmetric}) into
(\ref{eq:withpot}), we stumble upon some pleasing simplifications:
\ba
S_{\rm reg} &=& \int d^2x \, \sqrt{g_{(0)}} \,\Big\{
\epsilon^{-1} ( -\ft12 ) + 
\epsilon^{-1/2} (
 -\ft23 V_{(1)} ) + \log \epsilon (
-\ft18 R) + \CO(\epsilon^{1/2})  \Big\} \;.
\ea
{}From a pure supergravity point of view, the cancellations in the
$\log\epsilon$ term seem quite surprising: the back reaction of the
metric to the scalars exactly cancel the explicit contributions of the
scalars and leaves only the curvature.  Also in the GPPZ flow
\cite{BiFrSk01}, there is no new ``cross-term'' anomaly, even though
in that case as well as here, fluctuations of the active scalar and
metric are coupled.  In fact, holography provides a simple explanation
for this.  The boundary expectation value $\langle T^i_i \rangle$,
with scalar sources set to zero, is the trace of the variation of the
generating functional in the boundary field theory with respect to to
the boundary metric $g_{(0)ij}(x)$ --- 
also with scalar sources set to
zero.  Hence, gravitational and scalar anomalies on the boundary can
be separately computed, and must therefore simply add.  On the other
hand, one could have expected a separate scalar matter anomaly here
like the $\Phi \Box_{\gamma} 
\Phi + \ft16 R[\gamma] \Phi^2$ in four dimensions, 
but with $R[\gamma]=\epsilon R[g]$ and $\Box_{\gamma}=\epsilon
\Box_g$ we see that these terms will vanish
 as $\epsilon \rightarrow 0$ in our case.

We now proceed to regularize the action with local covariant
counterterms as outlined in \cite{BiFrSk01}. The philosophy is the
following: first, compute the divergent part of the regularized action
$S_{\rm reg}$ for a given potential. 
Second, pull back all quantities to the regulating
surface $\rho=\epsilon$, i.e. express the coefficients $q_{(n)}$ in
terms of $Q$, and analogously for the inert scalars.  
The full fields $Q(\rho,x)$ and $\Phi(\rho,x)$ are covariant, not
the individual coefficients in the $\rho$ expansion.  Third, the
covariant counterterm action $S_{\rm ct}(Q,\Phi)$ 
is defined by having the
same divergent parts as $S_{\rm reg}(Q,\Phi)$ already obtained, 
{\it and} causing the final renormalized action
\be
S_{\rm ren} = \lim_{\epsilon \rightarrow 0}
(S_{\rm
reg}+ S_{\rm ct})
\ee
to vanish on the background, which implies $\langle T_{ij}\rangle =0$.

We now perform these steps in the present setting.
For the inert scalars, substituting the $V_{(1)}$ given in (\ref{V1})
in the $\epsilon^{-1/2}$ term, we find
\be
S_{\rm reg} = \int d^2 x \sqrt{g_{(0)}}
\Big( -{1 \over 2 \epsilon} +{1 \over 4 \sqrt{\epsilon}}
(q_{(0)}^2 + \phi_{(0)}^2)
-\log \epsilon \; \ft18 R[g_{(0)}]  + \CO(\epsilon^{1/2}) \Big)
\label{eq:Sonshell}
\;.
\ee
To pull this back to $\rho=\epsilon$ we need
the inverse of \Ref{gexp}:
\be
\sqrt{g_{(0)}} = \sqrt{g}
\bigg\{1-\sqrt{\epsilon} \ft12 \Tr g_{(1)} + 
\epsilon[\ft14( \Tr g_{(1)})^2 -\ft12 \Tr g_{(2)} ]
-\epsilon \log \epsilon \ft12 \Tr h_{(2)} 
+ \CO(\epsilon^{3/2}) \bigg\} 
\; .
\label{eq:pullback}
\ee
Substituting (\ref{eq:pullback}) in (\ref{eq:Sonshell})
and again using \Ref{coeffmetric}, 
we find
\ba
S_{\rm reg} &=& \int d^2x \sqrt{g}
\; \Big(-{1 \over 2 \epsilon} -{1 \over 4 \sqrt{\epsilon}}
(q_{(0)}^2 + \phi_{(0)}^2) + 
\label{eq:Spullb}\\
&&  + \epsilon^0
\left (\ft18 (q_{(0)}^4 + \phi_{(0)}^4) 
-\ft34 \phi_{(0)}\phi_{(1)}+ \ft18 R[g_{(0)}]\right)
-\log \epsilon \; \ft18 R[g_{(0)}] + \CO(\epsilon^{1/2}) \Big)
\;.
\nn
\ea
(Notice the sign switch
of the $1/\sqrt{\epsilon}$ term.) If we would add noncovariant
counterterms to cancel only the divergences and then differentiate the
remaining finite action, as one essentially did in the ``old''
prescription, we would find the wrong one-point function. What is
wrong about it will be easier to see when we actually have the
one-point function at hand, e.g.\ \Ref{act1ptlin} for the active
scalar.

Instead, in holographic renormalization we concentrate on the
divergences in the covariant pulled-back action:
\be
S_{\rm reg} = \int d^2 x \sqrt{g}\,
\bigg( -{1 \over 2 \epsilon} - {1 \over \epsilon}
  (\ft14 \Phi^2+ \ft14 Q^2 )  
 -\log \epsilon \; \ft18 R[g] 
+ \CO(\epsilon^{0})   \bigg) \; .
\label{eq:Sctdetermine}
\ee
We remind the reader that it is $\phi_{(0)}$ and $g_{(0)}$ which are
kept fixed as $\epsilon \rightarrow 0$, hence e.g.\ 
$Q^2 \sim \epsilon^{1/2}$ as $\epsilon \rightarrow 0$.
Equation \Ref{eq:Sctdetermine} determines the divergent parts of the
covariant counterterm action $S_{\rm ct}$.  In total, including an
overall minus sign and using $\sqrt{g}R[g] = \sqrt{\gamma}R[\gamma]$, 
the full $2d$ counterterm action is
\ba
S_{\rm ct}
&=& 
 \int d^2 x \sqrt{\gamma} \; 
\bigg( \ft12  
+\ft14 Q^2+\ft14 \sum_{\bf i} (\Phi^\Si)^2  
+ a Q^4 + \sum_{\bf i} b_\Si\, Q^2 (\Phi^\Si)^2
+\log \epsilon \; \ft18 R[\gamma]    \bigg) \;, \non 
\label{eq:Sct}
\ea
where the index $\Si$ on $\Phi$ has been restored. 
We did not only cancel the divergences in
(\ref{eq:Sctdetermine}), we also added the smallest set of finite
counterterms needed for the renormalization scheme to preserve
supersymmetry; the coefficients $a$ and
$b_\Si$ are to be fixed. 

A few comments about the counterterms are in order.  First, the most
divergent term in the action is the volume divergence, which is
cancelled by a counterterm of $1/2$, or $(d-1)/2L$ in general $d$ and
for arbitrary AdS scale $L$.  This counterterm has been known since
the early days of the AdS/CFT correspondence, for general $d$ it was
introduced in \cite{LiuTse98}.  The quadratic counterterm $\ft14
(\Phi^{\Si})^2$ has also been familiar almost since the beginning.  In
\cite{BerSam01} the coefficient of this term was computed in the
fixed-background formalism for our case, and we indeed see it has the
same coefficient (in general, $(d-\Delta)/2$) as in the
fixed-background case, justifying that treatment in retrospect.  
A finite $Q^4$ counterterm was first displayed in \cite{BiFrSk01}, and
the $Q^2 (\Phi^i)^2$ terms can be seen as natural generalizations of
the $Q^4$ term, with the additional complication that the coefficients
$b^{\Si}$ cannot be determined by evaluation on the background, since
$\Phi_{\rm B}^{\Si}=0$. This is not a problem; in section
\ref{sec:susyWI} we will see how to compute $b^{\Si}$, using a
supersymmetry Ward identity for the two-point functions of
superpartner inert scalars.  The result is listed in
\Ref{tableb}.

 Notice we could not add trilinear terms like $Q
(\Phi^{\Si})^2$ since they would ruin the divergence structure; four
powers of $\epsilon^{1/4}$ scalars are needed to reach finiteness.
One could have added a finite counterterm proportional to the
curvature scalar $R$, but as is well known, $\int d^2x \sqrt{g} R$ is
a topological invariant in two dimensions (equal to $4\pi$ times the
Euler number) hence the variation of this term with respect to the
metric $g_{ij}$ is zero, so it would not contribute to correlators.

Finally, we fix the coefficient $a$ 
as in \cite{BiFrSk01} by evaluating 
$S_{\rm ren} = \lim_{\epsilon\rightarrow 0}
(S_{\rm reg} + S_{\rm ct})$ on the background: 
\ba
S_{\rm ren, B} &=& 
\int d^2 x \, \left(
aq_{(0)B}^4+\ft18 \tr (g_{(1)B}) q_{(0)B}^2 + \ft12 q_{(0)B}q_{(1)B}
+\ft14 \tr (g_{(2)B}) \right) \non
&=& \int d^2x \,\Big( a-{1 \over 96} \Big) \; , 
\label{a96}
\ea
where we used the background expansions \Ref{BGexpansion}.  Hence
$a=1/96$, and we see that the naive subtraction of finite counterterms
directly in the noncovariant action would have produced a different
result. There is a useful check of the counterterm coefficients
for the active scalar (see
e.g.~\cite{SkeTow99}): using the BPS equations, the expansion of the
negative of $W(Q)$ gives the coefficients directly, as $-W(Q)
=\ft12+\ft14 Q^2+\ft1{96}Q^4 + \ldots$ .

\subsection{One-point functions: generalities}

Now that we have the renormalized action, we can apply the formula at
the heart of the AdS/CFT correspondence in the supergravity
approximation:
\be
\underbrace{\langle\, e^{-\int d^2 x \; \sqrt{g}\, 
\left( q_{(0)} \CO_q + \phi_{(0)}^{\Si} \CO^{\Si}_{\phi}\right)  }
\,\rangle_{g_{(0)ij}}\rule[-3mm]{0mm}{3mm}}_
{\mbox{field theory}}
=\underbrace{ e^{-S_{\rm ren}
(Q,\Phi)}\rule[-3mm]{0mm}{3mm}}_{\mbox{supergravity}} \, ,
\label{eq:heart}
\ee
where the quantities on the left are the scaled Dirichlet data,
e.g.\ $q$ as opposed to $Q$. 

The variation of the renormalized action for small variations in the
sources $q_{(0)}$, $\phi_{(0)}$ and $g_{(0)}$ produces the boundary
one-point functions:
\be
\delta S_{\rm ren} 
= \int d^2 x \sqrt{g_{(0)}}\,
\Big( \ft12 \langle T_{ij} \rangle \delta g_{(0)}^{ij} 
+ \langle \CO_q \rangle \delta q_{(0)} 
+ \sum_\Si \langle \CO_{\phi}^{\Si} \rangle \delta \phi^{\Si}_{(0)} 
\Big) \; .
\label{eq:oneptdef}
\ee
Since bulk diffeomorphisms correspond to global symmetries of the
boundary theory, one should require these boundary quantities to
satisfy Ward identities.  Following \cite{BiFrSk01}, one can derive
Ward identities for the one-point functions of the stress-energy
tensor and scalars in the presence of sources:
\ba
\nabla^i \langle T_{ij} \rangle &=& -\langle \CO_q \rangle \nabla_j
q_{(0)}-\sum_\Si \langle \CO_{\phi}^{\Si} \rangle \nabla_j
\phi_{(0)}^{\Si} \;,   \label{eq:Ward1}   \\[0ex] 
\langle T_{i}^i \rangle &=& (\Delta_q-2)q_{(0)}\langle 
\CO_q \rangle +
\sum_\Si 
(\Delta_{\phi^{\Si}}-2)\,\phi_{(0)}^{\Si}\,
\langle \CO^{\Si} \rangle 
+ 
\CA  \non[0ex]
&=&  
-\ft12 q_{(0)}\,\langle \CO_q \rangle 
-\ft12 \sum_\Si \phi_{(0)}^{\Si}\,\langle \CO^{\Si} \rangle 
+ \CA \;,
\label{eq:Ward2} 
\ea
where $\CA$ is the conformal anomaly we saw arise 
in the previous section, from a
logarithmic counterterm breaking radial bulk diffeomorphisms.
If the reader finds this form of Ward identities unfamiliar,
it is probably because standard quantum field theory Ward identities
are usually expressed with sources set to zero, 
so that e.g.\ $\nabla^i \langle T_{ij} \rangle=0 $.

Explicitly, the anomaly $\CA$ arises
under a radial rescaling $\epsilon \rightarrow \mu^2
\epsilon$; all terms in $S_{\rm ren}$ above are manifestly invariant
except the logarithmic term.  It contributes an anomaly\footnote{The
factor $-\ft12$ is standard, cf. \cite{BiFrSk01} (5.43).} $-\ft12 \log
\mu^2 \CA$, hence we identify $\CA = -\ft14 R$.  This is as was to be
expected by holography; the conformal anomaly in 2d field theory on a
space of scalar curvature $R$ is simply proportional to $R$,
\[
- \ft14 R= -{L \over 2\kappa}R = -{c \over 24\pi}R = \langle T^i_i
\rangle  
\]
with $c=3 L /2 G_N$ the Brown-Henneaux central charge
and the gravitational coupling temporarily restored to
$\kappa=8\pi G_N$.  
The Ward identity (\ref{eq:Ward2}), including this anomaly, will
provide a useful check
for the one-point function of $T_{ij}$ computed below.

\subsection{Inert scalars}

The one-point function for the operators dual to inert scalars is now
easy to compute, there is one contribution from the regularized action
and some from the counterterms.  It is convenient to also introduce
the intermediate {\em subtracted} action $S_{\rm sub}(\epsilon) =
S_{\rm reg}(\epsilon) + S_{\rm
ct}(\epsilon)$, which is the quantity that becomes the renormalized
action $S_{\rm ren}$ when we take $\epsilon$ to zero: $S_{\rm ren} =
\lim_{\epsilon\rightarrow 0} S_{\rm sub}(\epsilon)$. 
The contribution from the 
regularized action \Ref{eq:Sreg} is
\bean
\delta S_{\rm reg} &=&
-2 \int_{\rho=\epsilon} \!\! d^2 x \sqrt{\gamma} \, 
\epsilon \sum_{\Si} \delta \Phi^{\Si} 
\partial_{\epsilon} \Phi^{\Si}
\eean
where we used the $\Phi$ bulk field equation
\Ref{eq:generalEOM} and
$\sqrt{g} = \epsilon \sqrt{\gamma}$.
Then, using the counterterm action 
$S_{\rm ct}$ from eq.\ (\ref{eq:Sct}) we can write down
the functional derivative
\bean
{1\over \sqrt{\gamma}}
{\delta S_{\rm sub} \over \delta \Phi^{\Si}} &=&
-2 \epsilon \partial_{\epsilon}\Phi^{\Si}
+\ft12  \Phi^{\Si}
+ 2b_{\Si} \, Q^2 \Phi^{\Si} 
 \\
&=& \epsilon^{1/4}\bigg[
-\ft12 \phi^{\Si}_{(0)}+\ft12 \phi^{\Si}_{(0)} 
+ \epsilon^{1/2}(-\ft32 \phi^{\Si}_{(1)} +\ft12\phi^{\Si}_{(1)}
- \tilde{\phi^{\Si}}_{(1)}
 + 2 b_{\Si}  q_{(0)}^2\phi^{\Si}_{(0)} )  \\
&& \qquad + 
\epsilon^{1/2} \log \epsilon (-\ft32 \tilde{\phi^{\Si}}_{(1)}+\ft12
\tilde{\phi^{\Si}}_{(1)})+ \ldots \bigg] \; .
\eean
This is divided by $\epsilon^{\Delta/2}=\epsilon^{3/4}$
and the limit $\epsilon\rightarrow 0$ is taken to yield 
the one-point function in the presence of sources
\ba
\langle \CO_{\phi^\Si} \rangle &=& 
-(\phi^\Si_{(1)}+\tilde{\phi}^\Si_{(1)})  
+ 2 b_\Si\, q_{(0)}^2\phi^\Si_{(0)} ~=~ -\phi^\Si_{(1)}  
+ 2 b_\Si \, q_{(0)}^2\phi^\Si_{(0)} \;,
\la{in1pt}
\ea
where we emphasize that this only 
holds to linear order in $\phi^{\Si}$, 
since we have only included $\Phi^{\Si}$ 
up to quadratic order in the 
action. The forefactor of $\phi_{(1)}$ is $-1$, or $-(2\Delta-d)$ in
general. (Notice that this is in \cite{BiFrSk01} conventions; there is
an overall sign switch in the one-point function from
\cite{dHSoSk00}). This forefactor also reproduces the one obtained in
\cite{BerSam01} using the fixed-background formalism, where 
the finite counterterm that appears here was neglected.

\subsection{Active scalar}

For the active scalar, the contribution of the regularized 
action \Ref{eq:Sreg} is
given by
\ba
\delta S = 
\int_{\rho=\epsilon} \!\! d^2 x \, \sqrt{\gamma} \:  \delta Q\Big(
-2 \epsilon  \, \partial_{\epsilon} Q - \epsilon 
\sum_\Si K_\Si \Phi^\Si \Phi^\Si  \partial_{\epsilon} Q 
\Big)
\; ,
\ea
where we used the bulk field equation for $Q$.
Proceeding as above, we obtain for the one-point function
\be
\langle \CO_q \rangle = -q_{(1)} + 4 a q_{(0)}^3 
+\, q_{(0)} \sum_\Si (2b_\Si-\ft14 K_\Si)\, \phi_{(0)}^\Si 
\phi_{(0)}^\Si  
\;.
\la{act1pt}
\ee
Expanding around the background \Ref{BGexpansion} according to
\ba
q_{(i)}  &=& q_{{\rm B}(i)} + \varphi_{(i)} \;,
\ea
we see explicitly that $\langle \CO_q \rangle$ vanishes on the
background for $a=1/96$ and there remain 
only the fluctuations:
\ba
\langle \CO_q \rangle &=& \ft18 \varphi_{(0)} - \varphi_{(1)} 
- \ft14\, \varphi_{(0)} 
\sum_\Si (2b_\Si-\ft14 K_\Si) \, \phi_{(0)}^\Si \phi_{(0)}^\Si  
+\CO(\varphi^2) \;.
\la{act1ptlin}
\ea
This confirms that the flow is a true operator deformation; the
boundary operator that drives the flow has vanishing vacuum
expectation value.  Notice that if we had not subtracted the finite
counterterm $aQ^4$ in the action,
or if we had tried to subtract a finite noncovariant counterterm
$a q_{(0)}^4$ directly, 
we would have obtained 
a nonvanishing result for the one-point function $\langle \CO_q
\rangle$; this would have been an apparent contradiction with the
claim that the deformation has vanishing vev. 

\subsection{Stress-energy tensor}
\label{sec:stressenergy}
We proceed to compute the stress-energy tensor one-point function in
the same way: functionally differentiate
the renormalized action with respect to the induced 
metric $\gamma^{ij}$ on the surface $\rho=\epsilon$,
but this computation is a little more complicated.
It is convenient to split the computation
into two parts, one corresponding to 
the extrinsic curvature, and one for the counterterms.
The gravitational contribution due to the extrinsic curvature term 
is\footnote{There is a sign mistake in \cite{BiFrSk01} 
(4.11). The correct expression is given in
\cite{dHSoSk00} (3.6).}
\bean
T_{{\rm reg, grav},\, ij} &=& {2 \over \sqrt{\gamma}}
{\delta \over \delta \gamma^{ij}} \:\Big( 
-\ft12 \int_{\rho =\epsilon}
d^2 x \, \sqrt{\gamma} \, \CK  \Big)
= -\ft12\,(\CK_{ij}-\CK\gamma_{ij}) \\
&=& -\ft12 (- \partial_{\epsilon} g_{ij}
+g_{ij}  g^{kl}\partial_{\epsilon}g_{kl} -{d-1 \over \epsilon}g_{ij} )
\eean
for general $d$, where by $\partial_{\epsilon}g_{ij}$
one intends $\partial_{\rho}g_{ij}(\rho,x)|_{\rho=\epsilon}$. 
Here the factor of 2 in the first expression
comes from (\ref{eq:oneptdef}), but it immediately
cancels with the 1/2 from the variation of $\sqrt{\gamma}$. 
For our metric ansatz we obtain
\bean
T_{{\rm reg, grav}, \, ij} &=&
-\ft12 \bigg[
\epsilon^{-1} (-g_{(0)ij}) 
+ \epsilon^{-1/2} (-\ft32 g_{(1)ij} +\ft12 g_{(0)ij} \trgo) \\ [2mm]
&& \quad
+\epsilon^0 (-2g_{(2)ij} - h_{(2)ij} + \ft12 g_{(1)ij} \trgo 
- \ft12 g_{(0)ij} \Tr g_{(1)}^2  + g_{(0)ij} \trgt) \\ [1mm]
&& \quad
+ \log \epsilon (g_{(0)ij} \trh -2h_{(2)ij}) + \CO(\epsilon^{1/2})
\bigg] \; ,
\eean
for the gravitational part.
The counterterm action contributes
\bean
T_{{\rm ct}, ij} &=& 
{2 \over \sqrt{\gamma}} {\delta S_{\rm ct} \over \delta \gamma^{ij}}
~=~  -\gamma_{ij} \: \Big(\ft12 
+\ft14 \sum_{\Si} (\Phi^{\Si})^2 +  
\ft14 Q^2 + a Q^4 + \sum_{\Si} b_{\Si} Q^2 (\Phi^{\Si})^2  \Big) 
\\ [1ex]
&=&
-\ft12\,\epsilon^{-1}\, g_{(0)ij}
- \ft12\,\epsilon^{-1/2} \,\Big[ g_{(1)ij} + g_{(0)ij}\,
\bigg( q_{(0)}^2 + 
\sum_{\Si}(\phi^{\Si}_{(0)})^2
\bigg) \Big] \\[2mm]
&&  -\epsilon^0\bigg[\ft12 g_{(2)ij} +
\ft14\, g_{(1)ij}\,
\bigg( q_{(0)}^2 + \sum_{\Si}(\phi^{\Si}_{(0)})^2 \bigg)
+\ft12\,g_{(0)ij}\,
\bigg( q_{(0)} q_{(1)} +   \sum_{\Si}
\phi^{\Si}_{(0)}\phi^{\Si}_{(1)} \bigg) \\ [1mm]
&& \hspace*{7cm}{}
+ g_{(0)ij}\,\bigg(a q_{(0)}^4 + 
\sum_{\Si} b_{\Si} \,q_{(0)}^2
(\phi^{\Si}_{(0)})^2  \bigg) \bigg] \\ 
&& \quad
-\ft12\,\log \epsilon\,  \Big[ h_{(2)ij} + 
g_{(0)ij} \,
\bigg(
q_{(0)}\tilde{q}_{(1)} + \sum_{\Si}
\phi^{\Si}_{(0)}\tilde{\phi^{\Si}}_{(1)} \bigg)  
+\CO(\epsilon^{1/2}) \Big] \;\;,
\eean
where the logarithmic term is scheme-dependent, being due to the
variation of any matter conformal anomaly.  Here, using the results of
sections \ref{sec:metriccoeff} and \ref{sec:scalarcoeff}, all terms at
order $\log \epsilon$ above actually vanish, but we will keep them
since the expressions may be useful in situations where they do not
vanish.  Also, the Ricci scalar $R$ does not contribute, as mentioned
above.

Putting everything together as $T_{\rm sub}=T_{\rm reg, grav}+T_{\rm
ct}$\,, the singular terms vanish upon using the earlier perturbative
expressions.  The finite part is the boundary one-point function, it
is given by
\ba
\langle T_{ij} \rangle
&=& \lim_{\epsilon\rightarrow 0} T_{{\rm sub},\,  ij} 
\la{T1pt}\\[1ex]
&=& \ft12 g_{(2)ij} +\ft12 h_{(2)ij} -\ft14 g_{(1)ij}
\Big[\trgo + (q_{(0)}^2 + \sum_\Si (\phi^\Si_{(0)})^2 )\Big] \non[1mm]
&& {}
+g_{(0)ij}\Big[\ft14 \Tr g_{(1)}^2-\ft12 \trgt 
-\ft12 q_{(1)}q_{(0)} - a q_{(0)}^4 
-\sum_\Si \ft12\,(\phi^{\Si}_{(1)} 
+ 2b_{\bf i}\, q_{(0)}^2\phi^\Si_{(0)} )\,\phi^\Si_{(0)} \Big]
\;.\nn
\ea
On the background this evaluates to
\ba
\langle T_{ij} \rangle_{\rm B} 
&=& \Big( {1 \over 96}-a \Big)\,\eta_{ij}  ~=~ 0 \;,
\nn
\ea
which vanishes by the value previously determined for $a$.
This is, of course, a trivial consequence of the 
fact that $a$ was defined to cause $S_{\rm ren}$ to vanish 
on the background.

Using \Ref{trg2} one may verify after some computation that \Ref{T1pt}
yields
\ba
\nabla^j \langle T_{ij} \rangle &=&
 \Big(q_{(1)} - 4aq_{(0)}^3 -\sum_\Si
(2b_\Si-\ft14 K_\Si)\, \phi^\Si_{(0)}\phi^\Si_{(0)} q_{(0)} \Big) 
\nabla_i q_{(0)}  \non
&&{}
+\sum_\Si
(\phi^\Si_{(1)} - 
2b_\Si\, q_{(0)}^2\phi^\Si_{(0)})\,\nabla_i\phi^\Si_{(0)}
\;,
\ea
and thus satisfies the Ward identity \Ref{eq:Ward1} with
the scalar one-point functions \Ref{in1pt}, \Ref{act1pt}. Similarly,
tracing \Ref{T1pt} and using the perturbative expressions for
$\trgo$ and $\trgt$, we obtain
\ba
\langle T_i^i \rangle &=& -\ft14 R[g_{(0)}] 
+ \ft12 \left(q_{(1)} - 4aq_{(0)}^3  \right) q_{(0)}
+ \sum_\Si
\ft12 \Big(\phi^\Si_{(1)} - 
(4b_\Si-\ft14 K_\Si)\, q_{(0)}^2\phi^\Si_{(0)}\Big)\,\phi^\Si_{(0)}
\;,
\nn
\ea
which together with \Ref{in1pt}, \Ref{act1pt}, guarantees the Ward
identity (\ref{eq:Ward2}) including the conformal anomaly $\CA =
-\ft14 R$.  Now we have our full collection of one-point functions and
proceed to compute two-point functions.

\section{Two-point functions of inert scalars}
\label{sec:twopoint}

\subsection{Fluctuation equations and the existence of prepotentials}
\label{sec:pre}

As has been emphasized above,
near-boundary analysis is no longer sufficient
when we move on to the computation of 2-point functions. It needs to be
supplemented with a solution to the equations of motion linearized around
the background \Ref{kink}. For the inert scalars, it follows from
\Ref{eq:scalard} that these fluctuation equations turn into a
three-dimensional Laplace equation in the domain wall metric, with
total potential
\ba
V^{\rm tot}_{\bf i} &=&
V_{\bf i} + \ft14\,K_{\bf i}\, (\dd_Q W)^2\Big|_{Q=Q_{\rm B}} \;.
\la{Vi0}
\ea
These total potentials were computed in~\cite{BerSam01}:
\ba
V^{\rm tot}_{\bf 1} &=& \ft1{1024}\,
( -45 - 160\,y + 10\,y^2 + 3\,y^4 ) \; ,
\non
V^{\rm tot}_{\bf 9{\rm +}} &=& - \ft1{16}\,
( 17 + 30\,y + y^2 ) \; ,
\non
V^{\rm tot}_{\bf 9{\rm -}} &=& \ft1{1024}\,
(y+1)( -93 + 13\,y - 19\,y^2 + 3\,y^3) \; , 
\non
V^{\rm tot}_{\bf 4{\rm +}} &=& -\ft1{16}\,
(3+y)(7+5y) \;,
\non
V^{\rm tot}_{\bf 4{\rm -}} &=& \ft1{1024}\,
(y+1)(y-5)(17 + 4\,y + 3\,y^2) \; ,
\la{Vi}
\ea
with $y=\cosh (\sqrt{2}Q_{\rm B})$. Further, it was shown
in~\cite{BerSam01} that the resulting equations of motion can be
transformed into one-dimensional Laplace equations in flat space,
with effective potentials $\CV_{\bf i}$ derived from
prepotentials (in the sense of supersymmetric quantum mechanics) as
$\CV_{\bf i} = \CU_{\bf i}' + \CU_{\bf i}^2$. This underlying
structure is crucial, as the absence of tachyonic fluctuations is then
manifest. Although it is easy to see that $\CV$ can always be
rewritten in terms of a prepotential $\CU$ for scalars with vanishing
explicit potential $V^{\rm tot}$ (i.e.\ when the effective potential
$\CV$ is only due to the $e^{2A}$ of the curved background,
cf.~\Ref{ViV} below), a priori it seems surprising that it would be
possible for all our inert scalars with the various potentials
above.  A similar situation was observed in~\cite{DWoFre00} for the
active scalar fluctuations in the most prominent exact
five-dimensional flows~\cite{BraSfe99,GPPZ00,FGPW00}, which seemed
somewhat puzzling and in need of explanation.

We now give a general argument for the existence of these
prepotentials in the fluctuation equations of gauged supergravity and
show how they may be directly extracted from the supergravity
Lagrangian.  Although we stay with our model as a concrete example,
the argument straightforwardly translates to other supergravities and
higher dimensions.  The fluctuations of inert scalars around the
background solution $\Phi^\Si$ are described by the bosonic Lagrangian
\ba
\CL^\Si = 
\ft12 \sqrt{G_{\rm B}}\,G_{\rm B}^{\mu \nu} 
\partial_{\mu} \Phi^\Si \partial_{\nu} \Phi^\Si  + 
\sqrt{G_{\rm B}}\,V^{\rm tot}_\Si(Q_{\rm B})\,\Phi^\Si\Phi^\Si
\;,
\ea
obtained from \Ref{eq:action} upon evaluation on the background. We
change to horospheric coordinates\footnote{Usually the horospheric
coordinate is called $z$, we call it $\zh$ to distinguish it from the
complex coordinate $z$ we introduce later.  }
\be
ds^2~=~e^{2A_{\rm B}(\zh)}\,(\eta_{ij}\,dx^i dx^j + d\zh^2) \;,
\qquad\quad\mbox{i.e.}\quad
\frac{d\zh}{dr}=e^{-A_{\rm B}} \; .
\la{conmet}
\ee
Redefining $\Phi^\Si=e^{-A_{\rm B}/2}R^\Si$ and dropping a total
derivative, this Lagrangian takes the form
\ba
\CL^\Si &=& 
\ft12\, \partial_{\mu} R^\Si \partial^{\mu} R^\Si  + 
\ft12\,\CV_\Si \,R^\Si R^\Si 
%-\ft14\dd^\mu\!\left(R^\Si R^\Si\dd_\mu A\right)
\la{Lflat}
\ea
in flat space, with a coordinate dependent potential 
\ba
\CV_\Si &=& 2e^{2A_B} V^{\rm tot}_\Si(Q_{\rm B}) +\ft12 A_{\rm
B}''(\zh)+\ft14 (A_{\rm B}'(\zh))^2 \;. 
\la{ViV}
\ea
Restoring the fermionic part of \Ref{L},
cf.~\cite{NicSam01b}, we arrive at the Lagrangian
\ba
\CL^\Si &=& 
\ft12\, \partial_{\mu} R^\Si \partial^{\mu} R^\Si  + 
\ft12\,\CV_\Si\,R^\Si R^\Si 
+\ft{1}{2}\,\Bchi{}^\Si \Gg^\mu\dd_\mu\chi^\Si 
+\ft12\, \CU_\Si\, \Bchi{}^\Si \chi^\Si 
\;,
\la{Lsusy}
\ea
which is invariant under the global $N=1$ supersymmetry
transformations 
\ba
\Gd R^\Si &=& \Bchi{}^\Si\Geps \;,\qquad
\Gd \chi^\Si ~=~  \Gg^\mu \Geps \, \dd_\mu R^\Si + 
\CU_\Si\,R^\Si\,\Geps \;,\qquad \Gg^{\zh}\Geps ~=~ \Geps \;,
\la{globalsusy}
\ea
that explicitly descend from the Killing spinors of the domain wall
solution \Ref{kink}. The fermionic mass term $\CU_\Si$ now serves as a
prepotential for the scalar potential
\ba
\CV_\Si &=& \CU_\Si' + \CU_\Si^2
\la{spqm}
\ea
and may be extracted from the corresponding mass term in the gauged
supergravity~\cite{NicSam01b}, more specifically from expanding the
so-called $A_3$ tensor around the background solution \Ref{kink} to
quadratic order of the inert scalars. The same prescription applies to
all higher-dimensional supergravities. This shows that the existence
of a prepotential is a direct consequence of the unbroken $N=1$
supersymmetry of the background.

Now, considering the full Lagrangian in the ${\bf 9}$ sector, say, it
is given by two copies of \Ref{Lsusy} with prepotentials $\CU_\Si$
related as $\CU_{{\bf 9}-}=-\CU_{{\bf 9}+}$. Closer inspection shows
that on the total system we can realize another supersymmetry ---
i.e.\ in addition to \Ref{globalsusy} --- with $\Geps$ of opposite
chirality $\Gg^{\zh}\Geps = -\Geps$, and where the supermultiplets are
$(R^{{\bf 9}+},\chi^{{\bf 9}-})$, $(R^{{\bf 9}-},\chi^{{\bf 9}+})$. In
other words, the fact that not only $N=1$ but $N=(1,1)$
supersymmetries are preserved implies that the potentials of the two
scalars $R^{{\bf 9}+}$, $R^{{\bf 9}-}$ are superpartners in the sense
of supersymmetric quantum mechanics, i.e.\ their prepotentials satisfy
$\CU_{{\bf 9}-}=-\CU_{{\bf 9}+}$. This in turn leads to the
correspondence between solutions of the equations of motion
\ba
R^{-} &=& (\dd_{\zh} -\CU_{+})\,R^{+} \;,
\la{R21}
\ea
which maps normalizable solutions into normalizable solutions.

Having understood where the prepotential structure comes from, let us
see how it can be exploited.  To facilitate later contact to standard
2d CFT expressions, we switch to a complex coordinate
$z=\frac1{\sqrt{2}}\,(x^1+ix^2)$ on the surfaces $\zeta=$ constant.
Then the plane wave ansatz is, using a complex variable also for the
2-momenta $p=\frac1{\sqrt{2}}\,(p^1+ip^2)$,
\be
R^{\Si}(\zeta,z) = e^{i(p\zb + \pb z)}R^{\Si}(\zeta)
\;,
\label{Ransatz}
\ee
and the fluctuation equations take the form
\be
(-\dd_{\zh}^2 + \CV_\Si\,) \,R^\Si = -2|p|^2 R^\Si \;,
\la{eomR}
\ee
with the coordinate $\zh$ from \Ref{conmet} and effective potentials
$\CV_{\bf i}$ derived from superpotentials~\cite{BerSam01}
\ba
\CU_{\bf 1} &=& \ft1{32}\, e^A\,(y-1)(y+11) 
\;,
\non
\CU_{\bf 9{\rm \pm}} &=& \mp\ft1{32}\, e^A\,(y-1)(y-5) 
\;,
\non
\CU_{\bf 4{\rm \pm}} &=&  \mp\ft1{32}\, e^A\,(y-1)(y+3) 
\;,
\la{Ui}
\ea
according to \Ref{spqm}.  Curiously, the scalar $\Phi^{{\bf 9}+}$ admits an
alternative superpotential $\tilde{\CU}_{\bf 9{\rm +}}=-\CU_{\bf 1}$,
a circumstance that is not explained by the above argument; carrying
different $SO(4)_{\rm inv}$ representations, $\Phi^{{\bf 9}+}$ and
$\Phi^{{\bf 1}}$ can of course not be in the same $(1,1)$
supermultiplet nor be related by another bulk symmetry.

Now, factoring out the zero mode ($p=0$) solution, and switching to a
new variable~$s$ related to~$y$ by $y=\ft{5s^3-2}{2+s^3}$, the general
solutions of the fluctuation equation \Ref{eomR} for $R^{\bf 9{\rm
+}}$ and $R^{\bf 4{\rm +}}$ may be written as
\ba
R^{\bf 9{\rm +}} &=& (1+y)^{-1/4}\, \chi_{-2} \;,\non
R^{\bf 4{\rm +}} &=& (1+y)^{1/12}(5-y)^{-1/3}\, \chi_{0} 
\;,
\la{Ri1}
\ea
with functions $\chi_\alpha(s)$ satisfying
\ba
s \,\chi_\alpha'' + (1+\alpha) \, 
\chi_\alpha'  -\frac{32 \,|p|^2}{3}\,
(2 + s^3 )\, \chi_\alpha &=& 0\;.
\la{ode}
\ea
The remaining solutions are then automatically obtained by putting the
supersymmetric quantum mechanics structure \Ref{R21} to work:
\ba
R^{\bf 1} &=& 
(\dd_{\zh} + \CU_{\bf 1})\,R^{\bf 9{\rm +}} \;, \non
R^{\bf 9{\rm -}} &=& 
(\dd_{\zh} + \CU_{\bf 9{\rm -}})\,R^{\bf 9{\rm +}} \;, \non
R^{\bf 4{\rm -}} &=& 
(\dd_{\zh} + \CU_{\bf 4{\rm -}})\,R^{\bf 4{\rm +}} \;.
\la{Ri2}
\ea
The two ordinary differential equations \Ref{ode} for $\alpha=0, -2$
thus comprise the entire dynamics of the inert scalar
fluctuations. Remarkably, we will see below that 
the fluctuation equations for
the active scalar and the metric reduce to the same
universal equation \Ref{ode}! Moreover, even the fluctuations in the
vector sector eventually lead to equations of type \Ref{ode}
\cite{BerSam01}.  This differential equation is a special case of the
{\em biconfluent Heun equation}, and we analyze it and its solutions
in appendix~\ref{AHeun}.

\subsection{Distinguishing $\Delta_+$ and $\Delta_-$}
\label{sec:Delta}

The conformal dimension $\Delta$ of the operator dual to a scalar
field of mass $m$ in two dimensions is
\[
\Delta = 1 \pm \sqrt{1 + m^2}  \; .
\]
As first pointed out in \cite{KleWit99}, there is a certain mass range
for scalars (here $-1 < m^2 \leq 0$) where the negative root can make
physical sense, hence one scalar can correspond to operators of two
different physical conformal dimensions $\Delta_{\pm}$.  Here we
summarize some recent arguments why the distinction between the two
roots is obvious for an active scalar, but is not for two inert
scalars belonging to conjugate roots $\Delta_{\pm}$ if they also have
the same quantum numbers.\footnote{Despite appearances, the ``$\pm$''
notation for e.g.\ $\phi^{{\bf 4}+}$ and $\phi^{{\bf 4}-}$ is not
intended to imply that $\phi^{{\bf 4}+}$ has to be dual to an operator
of dimension~$\Delta_+$; in fact, we will see that $\phi^{{\bf 4}+}$
is dual to an operator of dimension~$\Delta_-$.}

To begin, one can view the inert scalar expansions \Ref{expansions} as
being composed of two independent interlocking Taylor series,
beginning at orders $\rho^{(d-\Delta)/2}$ and $\rho^{\Delta/2}$,
respectively.  The former can be thought of as the ``source''
(corresponding to standard AdS/CFT usage) series and the other the
``response'' series, in the language of \cite{Muck01}. In the series
\Ref{expansions} for $\Delta=\Delta_+$, the exponent 1/4 is the
$(d-\Delta_+)/2$, whereas the ``response'' Taylor series starts at
order $\rho^{\Delta_+/2}=\rho^{3/4}$ and hence begins with the
$\phi_{(1)}$ and $q_{(1)}$ terms. As we have seen these terms are, in
fact, independent of the sources $\phi_{(0)}$ and $q_{(0)}$ in
near-boundary analysis (perturbation theory around $\rho=0$), but
acquire a ``nonperturbative'' interrelation upon demanding regularity of
the solution in the bulk interior.

For an operator with dimension given by the other root $\Delta_-$, in
our case $\Delta_-=1/2$, the ``response'' series begins already at
order $\Delta_-/2=1/4$, so essentially ``source'' and ``response'' are
interchanged. Now, if one attempts to compute the 2-point function in
the standard way, one is forced to add an additional quadratic
counterterm, as explained in \cite{Muck01}. For a field with
background (such as the active scalar $q$), this would generate terms
linear in the source, which yields a nonvanishing 1-point function even
when the source is set to zero. Hence, for the active scalar, selecting
$\Delta_-$ is distinguishable from selecting $\Delta_+$ in that the
flows describe quite different physics, and the former choice presumably
corresponds to turning on vevs in the boundary theory.\footnote{So
far, however, the machinery of holographic renormalization has only been
applied to flows whose active scalar is associated with an operator of
dimension $\Delta_+>d/2$.}

For an inert scalar, however, these linear terms are not
generated, and the only effect of the additional counterterm is that
the coefficient of $\phi_{(1)}\phi_{(0)}$ in the action switches sign
from $-(\Delta_+-d/2)$ to $(\Delta_+-d/2)= -(\Delta_--d/2)$.  
Thus, the flow and fluctuation equations are capable of
describing the correlation functions of operators of dimension
$\Delta_{-}$ equally well as $\Delta_{+}$.  This means that it is not a priori
obvious how to distinguish between inert scalars belonging to
different roots $\Delta_{\pm}$ if they also have the same quantum
numbers, in our case $SO(4)$ quantum numbers.  In the next section, we
will construct a supersymmetry Ward identity for two-point functions
including finite counterterms, which will, among other things, allow
us to make this distinction in section~\ref{sec:2ptinert}.

The 2-point functions for operators of dimensions $\Delta_+$ and
$\Delta_-$ are related by a ``massive'' Legendre transformation,
slightly generalized from~\cite{KleWit99} to include a
$\phi_{(0)}^{\Si}\phi_{(0)}^{\Si}$ term in the action with a
coordinate-independent coefficient $b_{\Si}$. As we have seen, such
terms arise as finite counterterms, but unlike the coefficient $a$ for
the active scalar, $b_{\Si}$ is not simply fixed by the domain wall
solution (cf.\ the discussion after \Ref{a96}).  In momentum space,
the renormalized action for the inert scalar $\phi^{\Si}$ can be
written as
\[
\hat{S}_{\rm ren}(\phi_{(0)}^{\Si}) =  \ft12 \int {d^4 p \over (2\pi)^2}
\phi_{(0)}^{\Si}(p) \phi_{(0)}^{\Si}(-p) 
(f_+(|p|) + b_{\Si}) \; ,
\]
where $f_+$ would have been
 the two-point function of operators of dimension $\Delta_+$,
had we neglected the finite counterterm.
The Legendre transform is effected by minimizing, just 
as in the massless case, the functional
\bean
J(\phi_{(0)}^{\Si},\phi_{(1)}^{\Si}) = 
\hat{S}_{\rm ren}(\phi_{(0)}^{\Si}) - \int {d^4 p \over (2\pi)^2}
\phi_{(0)}^{\Si}(p) \phi_{(1)}^{\Si}(-p) 
\eean
with respect to $\phi_{(0)}^{\Si}$, 
and in the linear approximation $\phi_{(1)}^{\Si}  = 
f_+(|p|)  \phi_{(0)}^{\Si}$. 
One solves for the extremum in $\phi_{(0)}^{\Si}$ and substitutes
in the original action to find
\[
\hat{S}_{\rm ren}(\phi_{(1)}^{\Si}) = -\ft12  \int {d^4 p \over (2\pi)^2}
\phi_{(1)}^{\Si}(p) \phi_{(1)}^{\Si}(-p) {1 \over f_+(|p|) 
+ b_{\Si}}  \; .
\]
In terms of the two-point function $f_-(|p|)$
of the Legendre-transformed (``response'') theory, this action should 
have the same sign as $\hat{S}(\phi_{(0)}^{\Si})$, in analogy to
the sign of the kinetic energy in transforming from Lagrangian to  
Hamiltonian for a free massive particle.
We can thus identify the two-point function of
the operator with dimension $\Delta_-$ as 
\be
\langle \CO_{\phi}^{\Si} \CO_{\phi}^{\Si} \rangle = 
-{1 \over f_+(|p|) + b_{\Si}} \; ,
\label{eq:Ominus}
\ee
if $f_+(|p|)$ is the would-be two-point function for the 
operator with dimension $\Delta_+$, when we neglect
finite counterterms. 
This expression will be useful in section \ref{sec:2ptinert}. 

\subsection{Counterterm fixed by Ward identity}
\label{sec:susyWI}

Now, what is this constant $b_{\Si}$ and how can it be computed?  To
answer this, we will derive the supersymmetry Ward identity between
correlation functions of inert scalars, embodying the supersymmetric
quantum mechanics structure.  In this subsection, let us again
suppress the representation index $\Si$ for clarity and only restore
it at the end.  Consider in general the correlation functions of
operators dual to a superpartner pair of scalar fields $\phi^{\pm}$.
That is, the scalars $R^{\pm}$ defined by
\ba
\phi^{\pm} &=& C(\rho) R^{\pm} 
\la{phiR}
\ea
for some function $C(\rho)$
(in our case $C(\rho)=e^{-A/2}$, but the explicit form is not
important for the moment) satisfy the two fluctuation equations
\ba
(-\dd_{\zh}^2 + \CV^{\pm})\,R^{\pm} &=& -2 |p|^2 R^{\pm} \;,
\la{flucsc}
\ea
with $\CV^{\pm}=\pm(\CU^{\pm})'+(\CU^{\pm})^2$.  The two
(normalizable) solutions of \Ref{flucsc} are then related by
\Ref{R21}.  Let us first focus on $R^+$, say.  Expanding $R^+$ in a
series in $\rho$,
\ba
R^+ &=& R_{(0)}^+ + \rho^{1/2} R_{(1)}^+ + \rho R_{(2)}^+ + \dots \; ,
\la{Rser}
\ea
the correlation functions of the dual operators 
in the complex coordinate $z = \frac{1}{\sqrt{2}}(x^1+ix^2)$ will be 
obtained in the next section
as (see \cite{BiFrSk01a})
\ba
\langle \CO^+_\phi(z)\CO^+_\phi(w) \rangle 
&=& 
-{1 \over \sqrt{g_{(0)}(w)}}{\delta \langle \CO^+_\phi (z)\rangle \over
\delta \phi_{(0)}(w)} \non [1ex]
&=& 
b + \frac{C_{(1)}}{C_{(0)}} + \frac{R_{(1)}^+(z)}{R_{(0)}^+(w)} \;.
\la{cfc}
\ea
where $C = C_{(0)} + \sqrt{\rho}\, C_{(1)} + \ldots$ as usual.
(Notice that in terms of the discussion of the previous section, this
is for a ``source'' series; for a $\Delta_-$ operator, the correlator
is actually the inverse of the right-hand side.)  In the remaining
part of this section we determine $b$.
 
{}From $d\zh/d\rho=(-2\rho \, e^A)^{-1}$ it follows that
\ba
\frac{\dd}{\dd {\zh}} &=& 
-2 \rho^{1/2}(1 + \xi \rho^{1/2} + \dots)\,\frac{\dd}{\dd \rho}
\;,
\label{eq:chain}
\ea
where the constant $\xi$ depends on the higher asymptotics in $A(\rho)$,
and is $\xi=-1$ in our case, but the precise value turns out not to be 
important. With the expansion \Ref{Rser}, we find
\ba
\dd_{\zh} R^+ &=& -R_{(1)}^+ - \rho^{1/2}\,(2R_{(2)}^+ + \xi
R_{(1)}^+) 
+ \dots 
\non[.5ex]
\dd^2_{\zh} R^+ &=& 2R_{(2)}^+ +\xi R_{(1)}^+ 
+ \dots ~\stackrel{!}{\equiv}~
R^+_{(0)} (\CV_{(0)}^+ + 2|p|^2) + \ldots \; .
\ea
Here the last equality uses the fluctuation equation \Ref{flucsc} and
the expansion 
\ba
\CV=\CV_{(0)} + \rho^{1/2}\CV_{(1)} + \dots \qquad
\Longrightarrow
 \quad \CV^+_{(0)} ~=~ -\CU_{(1)}^+ + (\CU_{(0)}^+)^2 \;.
\ea
The latter equation follows
from $\CV = \CU' + \CU^2$ and the chain rule \Ref{eq:chain}.
The expansion of the partner scalar $R^-$ 
is then already determined from \Ref{R21}:
\ba
R^- &=& (\dd_{\zh} -\CU^+)\,R^+ \non[1ex]
&=& -R^+_{(1)} - \rho^{1/2}\,R^+_{(0)} (\CV^+_{(0)}+2|p|^2) - 
(\CU_{(0)}^+ + \rho^{1/2}\CU_{(1)}^+)(R_{(0)}^+ +
 \rho^{1/2} R_{(1)}^+) + \dots
\non[1ex]
&=& -(R^+_{(1)} + \CU_{(0)}^+ R_{(0)}^+) - \rho^{1/2}\left(
\CU_{(0)}^+ R_{(1)}^+ + (\CU_{(0)}^+)^2 
R^+_{(0)} + 2 |p|^2 R^+_{(0)} \right) + \dots \; ,
\nn
\ea
from which we obtain the ratio
\ba
\frac{R^-_{(1)}}{R^-_{(0)}} &=& \CU_{(0)}^+ + 2\, |p|^2 \left(  
\CU_{(0)}^+  +
\frac{R^+_{(1)}}{R^+_{(0)}} \right)^{-1}
\ea
and using $\CU_{(0)}^+ = -\CU_{(0)}^-$, this
gives rise to the supersymmetry Ward identity
\ba
\left(\frac{R^-_{(1)}}{R^-_{(0)}}+\CU_{(0)}^- \right) \left(  
\frac{R^+_{(1)}}{R^+_{(0)}} + \CU_{(0)}^+ \right) &=& 2\, |p|^2 
\la{WI}
\ea
for coefficients $R_{(1)}/R_{(0)}$ describing\footnote{Again, for the
$\Delta_-$ operator the correlator itself is actually the inverse of
the expression in parenthesis, as we will see explicitly in the next
section. The Ward identity \Ref{WI} then
takes the familiar schematic form $\langle \CO \CO \rangle
=2|p|^2 \langle \CO \CO \rangle$. }  
two-point functions of two $N=(1,1)$ superpartner scalars.
Comparing to \Ref{cfc}, this determines the desired constant:
${b}^{\pm}$ is simply given by $\CU_{(0)}^{\pm}- C_{(1)} /C_{(0)}$,
i.e.\ $b^{\pm}$ is given by the boundary value of the
prepotential, up to a constant shift due to the 
curved background.  
Using the expansion \Ref{BGexpansion}, one finds
$C_{(1)}/C_{(0)}=1/4$ in this flow.  
We summarize the results as
\be
\begin{array}{|l|r|c|}\hline
\mbox{inert scalar} & \CU_{(0)}^{\Si} \quad & 
b_{\Si} =\CU_{(0)}^{\Si}-1/4  \\ \hline \hline
%\qquad  {\bf 1} & 3/8 \quad  & +1/8 \\  \hline
\qquad {\bf 9}+ & 1/8 \quad &  -1/8 \\  \hline
\qquad {\bf 9}- & -1/8 \quad & -3/8 \\  \hline
\qquad {\bf 4}+ & -1/8 \quad & -3/8 \\  \hline
\qquad {\bf 4}- & 1/8 \quad & -1/8 \\  \hline
\end{array}
\label{tableb}
\ee
These counterterm coefficients will be needed in the next section to
show that the proper behavior of two-point functions emerges without
dropping any terms by hand.

As one would expect, the argument 
leading to the supersymmetry Ward identity \Ref{WI}
can be seen in a different way on
a two-dimensional surface $\zh =$ constant,
using known results in (1,1) superspace. 
Substituting $\partial_{\zh} R^+ = \CU^+ R^+ + R^-$ in (\ref{Lsusy}), 
we find that for one chirality choice of the parameter
($\gamma^{\zh}\epsilon = -\epsilon$),
the supersymmetry transformation $\delta\chi_-$ has the superpartner
$R^-$  
precisely in the place where the top component $F$ 
of a (1,1) supermultiplet usually appears.
In other words, the component transformations of a real scalar 
(1,1) superfield $S$ with component fields $(R^+,\chi_+,\chi_-,R^-)$
precisely reproduce \Ref{globalsusy}.
Using standard results for relations between correlators
of different components of the same superfield,
one can finally recover a formula like \Ref{WI}.

\subsection{Two-point correlation functions}
\label{sec:2ptinert}

Since we study a flow to a fixed point, the boundary field theory is
conformal in both IR and UV limits. This means we expect asymptotic
power-law behavior of two-point functions (the only exception being
the ${\bf 4}$ marginal scalars in the IR, cf.~table~\ref{specL0}) on
both sides.  In other words, $\langle \CO_{\Delta} \CO_{\Delta}
\rangle \rightarrow p^{2\Delta-2}$ asymptotically, with two possibly
different values $\Delta_{\rm UV}$ and $\Delta_{\rm IR}$. If we would
apply the ``leading nonanalytic term'' (henceforth ``old'')
prescription for AdS/CFT correlators, as successfully applied in e.g.\
\cite{FMMR98}, we would quickly be disappointed in this case.  This is
because the correlator we are interested in is asymptotically just a
power of $p$, and in the ``old'' prescription it would have seemed
that the correct power-law behavior would be indistinguishable among
other monomial terms, coming from unphysical contact terms, that were
summarily dropped.

In fact, in holographic renormalization,
the requirement that only local counterterms may be
added is sufficiently restrictive 
to single out exactly the right behavior. Indeed,
$\sqrt{\gamma} \Phi \sqrt{\Box}_{\gamma} \Phi$
would yield a  counterterm $\sim p$ but is nonlocal,
whereas
$\sqrt{\gamma} \Phi \Box^n_{\gamma} \Phi$
would yield a $p^n$ counterterm but vanishes
in the limit $\epsilon \rightarrow 0$ for
all our $\Phi \sim \epsilon^{1/4}$ scalars.
In other words, holographic renormalization does not allow us
to ``drop contact terms'' as in the ``old'' prescription, which is
fortunate since precisely some of those monomial terms 
are physical ones in our context. (Of course, there are many
examples where that prescription is still applicable and useful.)

Now we compute the two-point functions of inert scalars.
These two-point functions descend
from uncoupled fluctuations that do not require an analysis of the kind
we will perform for the active scalar and metric in the next section.
They are obtained directly by taking the functional
derivative of the one-point function \Ref{in1pt}, and are 
thus 
essentially (i.e.\ up to finite counterterms) encoded in the
ratio $\phi_{(1)}^{\Si}/\phi_{(0)}^{\Si}$ in the expansion
\ba
\Phi^{\Si} &=& \rho^{1/4} \left( \phi_{(0)}^{\Si} + 
\sqrt{\rho}\,\phi^{\Si}_{(1)} + \dots \right) \;,
\ea
of the solution to the fluctuation equations, after we impose 
regularity in the
interior of the bulk. To compute this ratio, we recall that
$\Phi^\Si(\zeta,z) = e^{i (p \zb + \pb z)} \, e^{-A(\zh)/2} \,
R^\Si(\zh) $\,, and $R^\Si(\zh)$ is given explicitly in terms of
biconfluent Heun functions in \Ref{Ri1}, \Ref{Ri2}. Using the
following asymptotic expansions of the domain wall solution
\Ref{dwmetric}, \Ref{kink}
\ba
e^{-A/2} &=& \rho^{1/4}+\frac14\, \rho^{3/4} +
\frac{3}{64}\,\rho^{5/4} +\dots \; , 
\non 
y &=& 1+\rho^{1/2} + \frac{1}{4} \, \rho -\frac5{192} \rho^2 
+ \dots\; , 
\non 
%2^{1/4}\, (1+y)^{-1/4} &=& 1 -\frac1{8}\,\rho^{1/2} +
%\frac1{128}\,\rho + \dots \; , 
%\non 
s-s_0 &=& \frac1{4}\,\rho^{1/2} 
+ \frac1{16}\,\rho + 
\frac1{96}\,\rho^{3/2} +
\dots  \; .
\ea
(where we recall that $s$ is the variable used in the fluctuation
equation \Ref{ode}), we eventually find
\ba
{\phi_{(1)}^{\bf 1} \over \phi_{(0)}^{\bf 1} } &=& -\frac18
+ \frac{8|p|^2}{\Psi_{-2}(p)-2} 
\non  [1ex]
{\phi_{(1)}^{{\bf 9}+} \over \phi_{(0)}^{{\bf 9}+} }  
&=& \frac18 + \frac{\Psi_{-2}(p)}{4}
\qquad \; , \quad 
{\phi_{(1)}^{{\bf 9}-} \over \phi_{(0)}^{{\bf 9}-} }  ~=~ \frac38 + 
\frac{8|p|^2}{\Psi_{-2}(p)} 
\label{eq:inertratios} \\ [1ex]
{\phi_{(1)}^{{\bf 4}+} \over \phi_{(0)}^{{\bf 4}+} }  
&=& \frac38 + \frac{\Psi_0(p)}{4}
\qquad \; \, \quad
{\phi_{(1)}^{{\bf 4}-} \over \phi_{(0)}^{{\bf 4}-} }  ~=~ \frac18 + 
\frac{8|p|^2}{\Psi_{0}(p)} \nonumber
\ea
with the ratios $\Psi_{\alpha}$ from the expansion of the regular Heun
function in \Ref{expt0}. Consulting table \Ref{tableb}, we see that
the effect of the $Q^2\Phi^2$ counterterms with the coefficients
$b_{\Si}$ determined above is precisely to cancel the additive
constants in these expressions. The resulting correlators may be
expressed in terms of the Heun function coefficients
$\Psi_{\alpha}(p)$ 
(discussed in appendix \ref{AHeun}):
\ba
\langle \CO^{{\bf 1}}_\phi(-p) \CO^{{\bf 1}}_\phi(p) \rangle  
&=& k\, \frac{\Psi_{-2}(p)-2}{8\, (2+8|p|^2-\Psi_{-2}(p))}  \;,
\non [1ex]
\langle \CO^{{\bf 9}+}_\phi(-p) \CO^{{\bf 9}+}_\phi(p) \rangle  
&=& \frac{k}{2 \Psi_{-2}(p)} \;,
\qquad  \quad
\langle \CO^{{\bf 9}-}_\phi(-p) \CO^{{\bf 9}-}_\phi(p) \rangle  
~=~\frac{k |p|^2}{\Psi_{-2}(p)} 
\label{eq:inertcorrs} \;, \\ [1ex]
\langle \CO^{{\bf 4}+}_\phi(-p) \CO^{{\bf 4}+}_\phi(p) \rangle  
&=& \frac{k}{2 \Psi_0(p)} \;,
\qquad  \quad
\langle \CO^{{\bf 4}-}_\phi(-p) \CO^{{\bf 4}-}_\phi(p) \rangle  
~=~\frac{k |p|^2}{ \Psi_{0}(p)} \; , \nonumber
\ea
and all mixed two-point functions vanish.
Here we have restored the factor
$k/8$ which comes from properly normalizing
the supergravity action in \Ref{eq:heart}. This $k$ is the level
of the current algebra in
the ultraviolet CFT,  and it depends on 
the number $N$ of D5-branes as $k \sim N^2$ \cite{dBPaSk99}. 
We only restore this factor in final results
for correlators.

 The Heun function
coefficients $\Psi_{\alpha}(p)$ 
are plotted in figures~3 and~4 in appendix~\ref{AHeun};
analytic expressions for the large and small $p$ asymptotics are
obtained in \Ref{IR2}--\Ref{UV}, from which one may immediately derive
the UV and IR asymptotics of these correlators. Appealing to
\Ref{eq:Ominus}, we had to invert some of the ratios
\Ref{eq:inertratios}, namely those corresponding to the operators of
UV conformal dimension $\Delta_-=1/2$. In the ${\bf 4}$
and the ${\bf 9}$ sector, 
we want to make the distinction between $\Delta_+$ 
and $\Delta_-$. As it turns
out, this distinction is unique for the scalars in the ${\bf 4}$: it
is only upon inverting the shifted ratio $(\phi_{(1)}^{{\bf
4}+}/\phi_{(0)}^{{\bf 4}+}-3/8)$ that the resulting correlator has
the correct logarithmic behavior from~\Ref{IR0}, as expected for a
$\Delta_{\rm IR}=1$ operator. Tracing this identification back through
the flow, we conclude that in the UV it is $\Phi^{{\bf 4}+}$ which is
associated with $\Delta_-=1/2$, whereas $\Phi^{{\bf 4}+}$ corresponds
to $\Delta_+=3/2$. Thus, the computation of correlators throughout the
flow allows us to make this distinction which would have been
impossible to derive from a near boundary analysis around the UV
boundary.

In the ${\bf 9}$ sector, on the other hand, the 
ambiguity is left unresolved. We
have chosen to associate $\Phi^{{\bf 9}+}$ with $\Delta_-=1/2$
and $\Phi^{{\bf 9}-}$ with $\Delta_+=3/2$ but could also have done it
the other way round, both choices are compatible with the correct
asymptotics of $\Delta_-=1/2$, $\Delta_+=3/2$ operators in the UV and
IR. 

For the singlet $\Phi^{{\bf 1}}$, we know that it corresponds to a
$\Delta_-=1/2$ operator in the UV, since it is the superpartner of the
active scalar which has dimension $\Delta_+=3/2$ (see discussion
in section \ref{sec:Delta}). 
For this inert singlet scalar, we have fixed the coefficient of the
$Q^2\Phi^2$ counterterm such that the resulting correlator in
\Ref{eq:inertcorrs} satisfies the supersymmetry Ward identity with
the correlation function derived for the active scalar in
\Ref{eq:activecorrs} below. (It is quite a nontrivial
consistency check that this Ward identity may indeed be
satisfied by
just adding the proper constant to the ratio in \Ref{eq:inertratios}.)
This correlator leaves a minor puzzle in the IR: $\Psi_{-2}(p)$ goes
linear in $p$ for small $p$, so the correlation function goes to a
constant rather than like the $|p|^3$ which one would have expected
for $\Delta_{\rm IR}=5/2$.  However, irrelevant operators decouple in
the infrared, so it is difficult to know whether the method works
straightforwardly for operators crossing over from relevant to
irrelevant along the flow; certainly UV-irrelevant operators cannot be
treated the same way as UV-relevant operators
\cite{dHSoSk00}.

\section{Two-point functions of active scalar and stress-energy
tensor} 
\label{sec:active}

\subsection{Fluctuation equations}

Unlike the inert scalars, the active scalar fluctuation couples to the
metric fluctuation at linear order in the equation of motion.  This
was a tough obstacle (see e.g.\ \cite{DWoFre00}) until resolved in
\cite{BiFrSk01} (with earlier progress in \cite{ArFrTh00}). The
resolution involved working with ``gauge invariant quantities'',
therefore we vow to only work with such quantities, in a sense to be
made precise below.

To compute the quadratic fluctuations of the active scalar and the
metric, the inert scalars may be switched off as they do not
contribute to these couplings.  We use complex coordinates
$z=\frac1{\sqrt{2}}\,(x^1+ix^2)$ on the boundary, and parametrize the
domain wall background and the fluctuations as
\ba
ds^2 &=& e^{2A(r)} \Big(
2(1\pls h\pls |p|^2 H)\,  dz \, d\zb 
+\pb^2 (H\pls H_\bot)\, dz\, dz 
+p^2 (H\mis H_\bot)\, d\zb \, d\zb \Big) 
\non
&&{}
+ (1+h_{rr})\, dr^2 
\non
q &=& q_{\rm B} + \varphi
\la{flucact}
\ea
with the same plane wave ansatz in the complex $z$ coordinate as was
used earlier:
\ba
\varphi(z,\zb,r) 
= e^{i (p\zb + \pb z)}\,\varphi(r) \;,\qquad \mbox{etc.}
\ea
for all the fluctuations. Linearizing the equations of motion around
an arbitrary domain wall \Ref{dwmetric}, we obtain
\ba
H_\bot' &=& 0 \;,
\non
H''- 4 W H' &=& -e^{-2A} \,h_{rr} \;,
\non
2 h_{rr}\,W  + h'  &=&- 4 \varphi W' \;, \non
|p|^2 e^{-2A}\,h &=&  2|p|^2 \,W H'  +
(2 \varphi' - h_{rr} W' - 2 \varphi W'')\,W' \;,
\la{flucAM}
\ea
in terms of the superpotential $W(Q)$. In the above equations,
primes on the superpotential $W$
denote derivatives with respect to
$Q$, whereas all other primes refer to
derivatives with respect to the radial variable $r$. The first equation in
\Ref{flucAM} is a manifestation of the well-known fact that in three
spacetime dimensions there are no transverse-traceless degrees of
freedom in the metric.  As explained in \cite{BiFrSk01}, the ansatz
\Ref{flucact} does not completely fix the bulk diffeomorphisms, but
leaves the freedom of ``gauge transformations'' generated by vector
fields $(\xi^z,\xi^\zb,\xi^r)$ satisfying
\ba
\dd_r\xi^z = -pe^{-2A}\,\xi^r \;,
\quad \dd_r\xi^\zb = -\pb e^{-2A}\,\xi^r
\;,
\ea
under which the fluctuations transform as
\ba
\delta\, h &=& 2 \xi^r A' \;,\qquad
\delta\,h_{rr} ~=~ 2\dd_r\xi^r \;, \qquad
\delta\,\varphi ~=~ \xi^r q_{\rm B}' \;,
\non
\delta\,H &=& \ft1{|p|^2}\,(\pb \xi^z \pls p \xi^\zb) \;,\qquad
\delta\,H' ~= ~ -2 e^{-2A} \xi^r \;,
\non[.5ex]
\delta\,H_\bot &=& -\ft1{|p|^2}\,(\pb \xi^z \mis p \xi^\zb) \;,\qquad
\delta\,H_\bot' ~= ~ 0 \;.
\ea
It therefore seems appropriate to cast the above fluctuation equations
in equations for the ``gauge invariant'' objects
\ba
{\cal J}_1 &=& h + 4 \frac{W}{W'}\,\varphi \;,\qquad
{\cal J}_2 ~=~ -2 H' -\frac{4e^{-2A}}{W'}\,\varphi \;,
\non
{\cal R} &=& h_{rr} - \frac{2(\varphi' - \varphi W'')}{W'} \; .
\la{ginv}
\ea
This leads to
\ba
|p|^2\, e^{-2A}{\cal J}_1 &=& W {\cal R}' - 
\left(4 W^2  + (W')^2  - 2 W W''\right) {\cal R}  \;,
\non
|p|^2\, {\cal J}_2 &=& (-2 W'' + 4 W )\,{\cal R}  -  {\cal R}'  \;,
\non
{\cal J}_1' &=& -2\,W {\cal R}  \;,
\la{eomginv}
\ea
which we may combine into the second order equation for ${\cal R}$
\ba
0 &=&
{\cal R}'' + (2W''\mis 8 W)\,{\cal R}'  \non
&& \qquad  {}+\left( 2|p|^2 e^{-2A} + 16 W^2  \mis 4(W')^2
\mis 8 W W'' \pls 2 W' W''' \right){\cal R}
\;.
\la{eqR}
\ea
The analogous equation in five dimensions was obtained in
\cite{BiFrSk01}. Specializing this equation to the superpotential $W$
from \Ref{VQ} and transforming to  $y=\cosh(\sqrt{2}Q)$,
we find
\ba
2|p|^2\,e^{-2A}{\cal R}(y) &=& 
\ft1{64}(y-5)^2(y-1)^2(y+1)^2 {\cal R}''(y) \non
&&{}+
\ft1{64}(y-5)(y-1)(y+1)(11+5y^2) {\cal R}'(y) \non
&&{}+ \ft1{64}\left( 35 + y (80 + 10 y + 3
y^3) \right)\,{\cal R}(y)  \;.
\label{eq:REOM}
\ea
Dividing out its zero mode as in \Ref{Ri1},
\ba
{\cal R} &= & (y-1)(y-5)^{-\frac43}(y+1)^{-\frac23}\,\tilde{\chi} \;,
\ea
and performing the same change of variables
$y=\ft{5s^3-2}{2+s^3}$ as that leading to \Ref{ode},
eq.\ \Ref{eq:REOM} reduces to
\ba
s\,\tilde{\chi}'' - \frac{4+5s^3}{ 2 + s^3}\,
\tilde{\chi}' 
- \frac{32\,|p|^2}{3}\, (2 + s^3) \,\tilde{\chi} &=& 0 \; .
\la{odem}
\ea
We now proceed to show that also this equation may be reduced to one
of the two Heun equations~\Ref{ode}.  Indeed, let $\chi$ be the
regular solution of \Ref{ode} with $\alpha=-2$, then
\ba
\tilde{\chi} &=& \chi - \ft12\,s\chi'
\ea
is a solution of \Ref{odem}, i.e.\ we can extract all the asymptotics
data from our previously obtained solution! In particular, after some
computation it is seen that the solution of \Ref{eqR} regular
in the bulk interior has the expansion
\ba
{\cal R} &=& {\rm const} \times \left( \sqrt{\rho} + 
\frac{-2 + 32 |p|^2 + \Psi_{-2}(p)}
{4 (-2 + \Psi_{-2}(p))}\,\rho + \dots \right) \;,
\la{ratioR}
\ea
with $\Psi_{-2}(p)$ from \Ref{expt0}. Using this expansion we can now
procced to compute two-point functions.

\subsection{Two-point correlation functions}
\label{sec:2ptactive}

The rest of the computation is straightforward.
Linearizing the one-point function of the stress-energy tensor
\Ref{T1pt} around the background using
\ba
g_{ij} &=& \rho e^{2A_{\rm B}} (\eta_{ij}+ h_{ij}) \;,
\ea
leads to
\ba
\langle T_{ij} \rangle
&=& \eta_{ij} \left(
\ft14 \Tr h_{(1)}- \ft12 \Tr h_{(2)} 
+\ft7{16} \varphi_{(0)} -\ft1{2} \varphi_{(1)} \right) 
- \ft14 h_{(1)ij}+ \ft12 h_{(2)ij} \;,
\la{T1ptlin}
\ea
or, in the complex notation from \Ref{flucact},
\ba
\langle T_{z\bar{z}} \rangle
&=& 
\ft14 ( h_{(1)} + |p|^2 H_{(1)}) 
-   \ft12 ( h_{(2)} + |p|^2 H_{(2)})
+\ft7{16} \varphi_{(0)} -  \ft12 \varphi_{(1)}  \; , \non
\langle T_{zz} \rangle
&=&  \bar{p}^2 \left(
- \ft14 H_{(1)}  + \ft12 H_{(2)} \right)
\; , \non
\langle T_{\bar{z}\bar{z}} \rangle
&=&  p^2 \left(
- \ft14 H_{(1)} + \ft12 H_{(2)} \right) \; .
\ea
where the subscripts in parentheses 
denote coefficients in the $\rho$ expansion,
$H=H_{(0)} + \sqrt{\rho}H_{(1)} + \rho H_{(2)} +\ldots $ as
usual, and we have used that $H_\bot=$ constant.  Expanding the gauge
invariant quantities \Ref{ginv} yields
\ba
{\cal R}  &\equiv& {\cal R}_{(0)} \sqrt{\rho} 
+ {\cal R}_{(1)} \rho + \dots  \non
&= &
\left(\ft12 \varphi_{(0)} - 4 \varphi_{(1)} \right) \sqrt{\rho} +
\left(-\ft{15}{16}\,\varphi_{(0)} + \varphi_{(1)} - 8 \varphi_{(2)} 
\right) \rho +
\dots \; , 
\non [1ex]
{\cal J}_{(1)} &=& (h_{(0)}+4 \varphi_{(0)}) + \left(h_{(1)} + \ft32 
\varphi_{(0)} + 4 \varphi_{(1)}
\right) 
\sqrt{\rho} \non
&& \qquad + \left( h_{(2)} + \ft{15}{32}\,\varphi_{(0)} + \ft32\,
\varphi_{(1)} +
4 \varphi_{(2)} 
\right) \rho + \dots \; ,  
\non [1ex]
{\cal J}_{(2)} &=& 2H_{(1)} \sqrt{\rho} + \left(4H_{(2)}+ 8 
\varphi_{(0)}\right) \rho + \dots \; .
\ea
One sees that 
 the coefficient ${\cal R}_{(0)}=\ft12\varphi_{(0)}-4\varphi_{(1)}$ 
is proportional to
the one-point function $\langle O_q \rangle $
of the active scalar 
(just set $\phi^{\Si}= 0$ in 
\Ref{act1ptlin}), hence that the
one-point function is gauge invariant
as one would have hoped.  From the equations of motion
\Ref{eomginv}, we then find that we can express all the perturbative
coefficients in terms of $\varphi_{(0)}$, $h_{(0)}$, and the ratio
${\cal R}_{(1)}/{\cal R}_{(0)}$ which has been determined in
\Ref{ratioR} above:
\ba
H_{(1)} &=& 0 \;,\qquad 
H_{(2)} ~=~ \ft12\,h_{(0)} + \frac{h_{(0)} + 4 \varphi_{(0)}}
{4({\cal R}_{(1)}/{\cal R}_{(0)})-5} 
\;,
\non
h_{(1)} &=& -2\,\varphi_{(0)} \;,\qquad 
h_{(2)} ~=~ -\ft14 \,\varphi_{(0)}
+ \frac{2|p|^2(h_{(0)} + 
4 \varphi_{(0)})}{4({\cal R}_{(1)}/{\cal R}_{(0)})-5}
\;,
\non
\varphi_{(1)} &=& \ft18 \,\varphi_{(0)}
- \frac{2|p|^2(h_{(0)} + 
4 \varphi_{(0)})}{4({\cal R}_{(1)}/{\cal R}_{(0)})-5}
\; .
\ea
Substituting this into \Ref{act1ptlin}, \Ref{T1ptlin}, we find
\ba
\langle \CO_q \rangle &=& 
\frac{2|p|^2\,(h_{(0)} + 
4 \varphi_{(0)})}{4({\cal R}_{(1)}/{\cal R}_{(0)})-5} \;,
\non[1ex]
-\frac1{|p|^2}\, \langle T_{z\bar{z}} \rangle
&=& \frac1{\bar{p}^2}\,\langle T_{zz} \rangle ~=~
\frac1{p^2}\,\langle T_{\bar{z}\bar{z}} \rangle ~=~
\ft14\,h_{(0)} + \frac{h_{(0)} + 4 \varphi_{(0)}}{8({\cal
R}_{(1)}/{\cal R}_{(0)})-10} \;, 
\ea
which gives the two-point correlation functions
\ba
\langle \CO_q\CO_q \rangle &=& 
\frac{k|p|^2}{4({\cal R}_{(1)}/{\cal R}_{(0)})-5}~=~ 
\frac{k |p|^2 \,(-2 + \Psi_{-2}(p))}{4(2 + 8 |p|^2 - \Psi_{-2}(p))}
\;,
\non [1ex]
\langle \CO_q T_{z\bar{z}}\rangle &=& 
\frac{-k|p|^2}{16{\cal R}_{(1)}/{\cal R}_{(0)}-20} ~=~ 
-k|p|^2\, 
\frac{-1 + \ft12 \Psi_{-2}(p)}{8(2 + 8|p|^2 - \Psi_{-2}(p))}
\;,
\la{eq:activecorrs}
\\ [1ex]
\langle T_{z\bar{z}}T_{z\bar{z}}\rangle &=& -\frac{k}{32}\,|p|^2 \,
\frac{4{\cal R}_{(1)}/{\cal R}_{(0)} -3}
{4{\cal R}_{(1)}/{\cal R}_{(0)} -5} ~=~ 
- {k|p|^4 \over 8}\left(\frac1{8|p|^2} +\frac{1}{2+8|p|^2 - 
\Psi_{-2}(p)}\right)
\;, \nn
\ea
where we have substituted the expansion \Ref{ratioR} for ${\cal
R}_{(1)}/{\cal R}_{(0)}$.  
The first two correlators are, now that the
smoke has cleared, trivially related by\footnote{see e.g.\
\cite{Sken02} eq.\ (6.13), and use $\langle \CA \CO_q \rangle_{\rm B}
= 0$.}
\be
T_{z \zb} = \beta \CO_q + \CA \;,
\ee
where $\beta=-1/2$ is the classical $\beta$ function due to the
classical scaling of the coefficient $q_{(0)}$, when this coefficient
is viewed as the holographic coupling constant in the deformation
$\CL_{\rm CFT} + q_{(0)} \CO_q$.  As in \cite{BiFrSk01}, 
the fact that the $\beta$ function is classical could be 
ascribed to a 
nonrenormalization theorem to the effect that the only contribution to
the $\beta$ function could come from an anomalous dimension of the operator
it multiplies, and if this operator is protected, there are no quantum
corrections to scaling. After all, the deformation preserves some
supersymmetry, so this is perhaps not so surprising.

It is straightforward to see that the two-point correlator of the
active scalar has the correct (linear $|p|$) UV behavior,
corresponding to an operator of conformal dimension $3/2$. Moreover,
comparing to \Ref{eq:inertcorrs}, we see that it indeed satisfies the
supersymmetry Ward identity $\langle \CO_q\CO_q \rangle = 2|p|^2
\langle \CO_\phi^{\bf 1}\CO_\phi^{\bf 1} \rangle$. The third correlator,
$\langle T_{z\bar{z}}T_{z\bar{z}}\rangle$, and its asymptotics will be
studied in more detail in the next section, since it is related to the
$C$ function along the renormalization group flow.

\mathon
\section{The $C$ function}
\label{sec:C}
\mathoff

Given the stress-energy 2-point functions, we may follow
Zamolodchikov's original construction \cite{Zamo86} to compute 
the {\it $C$ function}, a function on the space of couplings that is
monotonic along the flow and interpolates between the central charges
of the conformal fixed points.
The variety of possible
tensor structures in $d>2$ makes this construction difficult 
to generalize to higher dimensions 
\cite{CaFrLa91,Anse97}; in particular, the
straightforward proof of monotonicity as a consequence of unitarity is
tailored to fit the two-dimensional case. Nevertheless, there have been several
proposals for defining monotonic $C$~functions by holography
\cite{AlvGom98,GPPZ98,FGPW99}. In particular the ``holographic $C$
function'' of \cite{GPPZ98,FGPW99} is a very simple proposal in terms
of the supergravity superpotential $W$, which reads
$C_{\rm hol}\equiv-1/{W(Q_{\rm B})}$ in two dimensions. It
is monotonic as a function of the bulk radial coordinate, assuming a
fairly weak positive-energy condition in the bulk
supergravity. Applied to our superpotential \Ref{VQ}, we find
\ba
C_{\rm hol} &=& -\frac{3k}{W(Q_{\rm B})} ~=~ 
3k \, \Big(\, 1 - \frac12\,\sqrt{\rho} +
\frac{3}{16}\,\rho + \dots \Big) \;.  \la{Chol} 
\ea 
with normalization adapted to our conventions.
Monotonicity may be verified directly.

\begin{figure}[htbp]
  \begin{center}
\epsfxsize=85mm
\epsfysize=55mm
\epsfbox{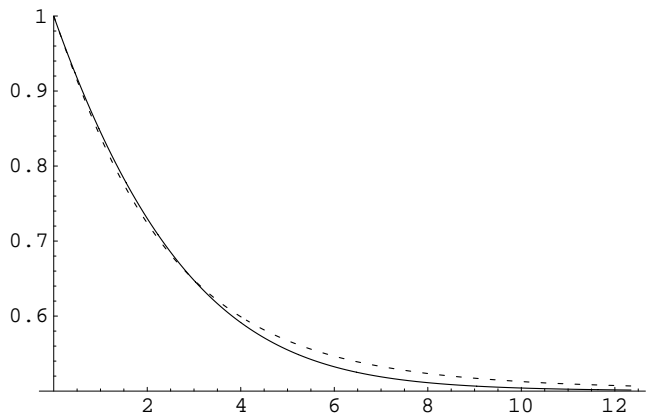}
  \caption{\small Holographic $C$ function \Ref{Chol} (solid) and
  Zamolodchikov's $C$ function \Ref{CZam} (dashed) from holographic
  correlators as functions of $\mu$ after normalizing and rescaling
such as to match the first derivative in the UV. 
} \label{Cpic} \end{center}
\end{figure}

We will now compute Zamolodchikov's $C$ function from the
holographic $\langle TT \rangle$ correlators obtained above.
As a first check of the $\langle TT \rangle$ correlators
\Ref{eq:activecorrs}, we expand their asymptotic behavior using
\Ref{IR2}, \Ref{UV}, and find
\ba
\left\langle T_{z\bar{z}}(p)T_{z\bar{z}}(-p)\right\rangle 
&\stackrel{{\rm UV}}{=}& 
-{k \over 32}\:\Big(\,
 |p|^2 - \frac{1}{2\sqrt{2}}\,|p| + \frac{5}{32} + 
\dots \Big)\; ,
\non
\non
\left\langle T_{z\bar{z}}(p)T_{z\bar{z}}(-p)\right\rangle 
&\stackrel{{\rm IR}}{=}& 
-{k \over 32}\:\Big(\,
 \frac12\,|p|^2 + 2 |p|^4 -\frac{4\sqrt{2}}{\sqrt{3}}\,|p|^5
+  \dots \Big)\; .
\la{TTasym}
\ea
In particular, this shows that in fact $c_{\rm IR}/c_{UV} = 1/2$,
which is precisely what is expected (see eq.\ \Ref{ccVV}). To construct
Zamolodchikov's $C$ function, we first Fourier
transform the $\left\langle
T_{z\bar{z}}T_{z\bar{z}}\right\rangle$ correlator back to the complex
$z$ plane as
\ba
\left\langle T_{z\bar{z}}(z)T_{z\bar{z}}(0)\right\rangle 
&=&
\frac1{2\pi}\,\int dp \, d\bar{p}
\, e^{i(p\bar{z}+\bar{p}z)}\,
\left\langle T_{z\bar{z}}(p)T_{z\bar{z}}(-p)\right\rangle 
 \non
&=: &
\square\,\square\, \Omega(t) 
=
\frac4{|z|^4}\,(\Omega''-\Omega'''+\ft14\Omega'''') 
\;,
\ea
defining the function $\Omega(t)$ with $t=\ft12\log (\mu^2 |z|^2)$,
that is, $\dd_t=|z|\dd_{|z|}$, and primes denote derivatives
with respect to $t$. 
This function encodes Zamolodchikov's $C$
function \cite{Zamo86} as
\ba
C_{\rm Zam} &=& -96\, ( \Omega'-\Omega''+\ft14\Omega''' ) \; ,
\la{CZam}
\ea
such that $C'=-24\,|z|^4\,\left\langle
T_{z\bar{z}}(x)T_{z\bar{z}}(0)\right\rangle $. Specifically we find
the integral representations
\ba
\Omega(t) &=& \int d|p| \, |p|^{-3}\,J_0(2|pz|) \,
\left\langle T_{z\bar{z}}(p)T_{z\bar{z}}(-p)\right\rangle \; ,
\non
\Omega'(t) &=& -\int d|p|\,  2|p|^{-2}\,J_1(2|pz|) \,
\left\langle T_{z\bar{z}}(p)T_{z\bar{z}}(-p)\right\rangle\; ,
\non
C_{\rm Zam} &=& 192
\int d|p| \Big\{
|p|^{-2} |z| \left( |pz|J_0(2|pz|) +
(|pz|^2-1)J_1(2|pz|) \right) \non[-1ex]
&& \hspace{8cm} \times 
\left\langle T_{z\bar{z}}(p)T_{z\bar{z}}(-p)\right\rangle
\Big\} \; . \la{CFT}
\ea
With proper regularization, the last integral gives
\ba
\int d|p| \Big\{
|p|^{-2} |z| \Big( |pz|J_0(2|pz|) + \hspace{3cm} && \\ [-2ex]
 (|pz|^2-1)J_1(2|pz|) \Big) \,
|p|^{n}
\Big\} &=& 
\frac{\pi\,n\,|z|^{2-n}}
{4\sin(\frac{n\pi}{2})\Gamma(2-\frac{n}2)\Gamma(-\frac{n}{2})}\; ,
\nn
\ea
for the polynomial terms in $\left\langle
T_{z\bar{z}}T_{z\bar{z}}\right\rangle$. From the exact asymptotics
\Ref{TTasym}, we may then derive the small distance behavior of the
$C$ function
\ba
C_{\rm Zam} &=& 3k  \left( 1 - \frac1{4\sqrt{2}} \,|z| + \CO(|z|^3)
\right) \;.
\ea
For the full $C$ function we have to insert in \Ref{CFT} the complete
expression from \Ref{eq:activecorrs}, with $\Psi_{-2}$ from appendix
$B$. The numerical result is
 plotted in figure~\ref{Cpic}, and it is another
consistency check of the correlators \Ref{eq:activecorrs} that 
the function
comes out strictly monotonic, i.e.\ the holographic correlators indeed
reproduce the properties of a unitary field theory. 

To compare the result to the proposal \Ref{Chol} above, we recall
that in holographic flows the RG scale $\mu$ is introduced by the AdS
isometry
\ba
\rho&\ra&\mu^2\rho \;,\qquad x^i~\ra~ \mu x^i \;,
\label{eq:scalingiso}
\ea
connecting the energy scale to the bulk radial coordinate. At
fixed $\rho$ this converts the superpotential $W(Q_{\rm
B}(\rho=\mu^2\rho_0))$ into a function of $\mu$. Likewise, $C_{\rm
Zam}$ turns into a function of $\mu$ upon relating the boundary radius
$|z|$ to a fixed value $|z|=\mu|z_0|$. Normalizing $C_{\rm hol}$ and
$C_{\rm Zam}$ to unity in the UV and fixing the values of $|z_0|$ and
$\rho_0$ such that the first derivatives $C_{\rm hol}'$ and $C_{\rm
Zam}'$ coincide in the UV, we plot the two functions in
figure~\ref{Cpic}. 

Both functions are strictly monotonic with surprisingly similar
shapes. We should stress that the discrepancy in figure 
\ref{Cpic} does not
mean that the different proposals are incompatible --- the computation
of the $C$ function is in any case scheme dependent --- but may rather
indicate that the identification of the boundary energy scale with the
radial AdS variable \Ref{eq:scalingiso} requires corrections away from
criticality.

A comparison of the different $C$ function proposals in the 
five-dimensional confining flow of \cite{GPPZ00} was done in
\cite{AGPZ00}, see also \cite{Erdm01}. 
Some general considerations in higher-dimensional
conformal-to-conformal flows are presented in
\cite{MarMie01}. Note that in higher dimensions the relevant central
function descends from the two-point correlators of the
transverse-traceless part of the stress-energy tensor, which
drastically simplifies the computation. In two dimensions one instead 
has to go through the full procedure of decoupling the active scalar
and metric fluctuations, leading to the fluctuation equation \Ref{eqR}
and its solution presented in the previous section.

A challenging goal would of course be the computation of the $\langle
TT \rangle$ correlators directly in the CFT as input to $C_{\rm
Zam}$. Comparison to the correlators obtained by the holographic
methods presented here could provide a demanding test of the
correspondence.

\section{Outlook}

In this paper, we have computed correlators along a renormalization
group flow interpolating between conformal field theories. As a main
result, we have obtained the two-point correlation functions
$\langle TT \rangle$, $\langle T \CO \rangle$, and $\langle \CO\CO
\rangle$ of the stress-energy tensor, and operators $\CO$ dual to
supergravity scalars, both active and inert. We used the $\langle TT
\rangle$ correlators to compute the Zamolodchikov $C$ function and
compare it to the holographic proposal for the $C$ function in terms
of the supergravity superpotential.  Since we summarized the results
in the introduction, here we only give some brief remarks on future
directions.

First, one could study correlators of currents $J^i$ in the boundary
theory. As in \cite{BiFrSk01a}, these currents correspond to
symmetries broken by the deformation $q_{(0)}\CO_q$ (here $SO(4)$),
although their analysis was facilitated by the remaining $U(1)_R$
symmetry of the IR theory, whereas our IR theory has no $R$-symmetry
at all.  Presumably this analysis is nevertheless straightforward on
the supergravity side, given the fact that also the vector fluctuation
equations reduce to the same
biconfluent Heun equation \cite{BerSam01}.

More interestingly, it would be worth to study the CFT side in more
detail than has been done here. In fact, it should be possible to
compute correlators such as $\langle TT \rangle$ given the deformation
$\CL_{\rm CFT} + q_{(0)}\CO_q$, where $q_{(0)}$ is the holographic
coupling. One could hope to show that this deformation is integrable,
and to compute correlators exactly.

One can also proceed to apply these methods in other settings.
Perhaps the example of greatest direct physical interest would be
domain wall solutions in five-dimensional gauged supergravity.  Even
though the 5d flow equations in \cite{FGPW99} could not be solved
exactly, one could see how far one can take a numerical analysis of
correlators there. Unfortunately, as emphasized in section
\ref{sec:summary}, it will probably be well-nigh impossible to achieve
any confidence in the numerics if one cannot find any analytic help.
On the other hand, it might be possible 
to construct exactly soluble setups in
other five-dimensional cases than that of~\cite{FGPW99}. Encouraged by
the solution in section \ref{sec:summary}, and recalling that it was
crucial for the explicit construction of \cite{BerSam01} to set the
two D5-brane charges equal (other values of this ratio would have
resulted in a vertical stretching of the potential in
figure~\ref{Vpic} such that the flow trajectory would no longer be a
straight line), one might look for simplifying special values of
coupling constants in various five-dimensional gauged
supergravities. In an analogous half-maximal theory in 5d, in which
first-order equations for flows were studied in \cite{CGWZ02}, there
is an $SU(2)_L \times SU(2)_R$ non-Abelian gauge symmetry but
apparently no free parameter of this kind.

As a matter of principle, it would be interesting to compute $n$-point
functions for $n >2$.  Although the full Dirichlet problem may be
practically unsoluble, one can imagine solving nonlinear fluctuation
equations in perturbation theory.  Recently, an outline of an example
for $n=4$ was given in \cite{Sken02}, section 5.9.

Finally, one could compute correlators in other RG flows in
two-dimensional conformal field theories, further pursuing the idea of
three-dimensional gauged supergravity as a tool in this field. Of
particular interest and in principle accessible with our tools are for
example flows to nonsupersymmetric but stable fixpoints, the analysis
of marginal deformations of the CFT describing the D1-D5 system
(cf.~\cite{Dijk98,DaMaWa99,AhBeSi01}), and flows in the maximally
supersymmetric theory of \cite{NicSam00} and their role in a
supergravity description of matrix string theory
(cf.~\cite{MorSam02}).

\section*{Acknowledgements}

We gratefully acknowledge helpful discussions with 
B.\ Farid,
I.\ Runkel and
K.\ Skenderis,
 and we
especially thank M.\ Bianchi and M.\ Haack for 
useful discussion and comments on early
drafts of this paper. 
We thank IHES, Bures-sur-Yvette,
for hospitality when this work was being finalized.
M.B. is supported by a Marie Curie Fellowship,
contract number HPMF-CT-2001-01311.
This work was supported in part by
INFN, by the EC contract HPRN-CT-2000-00122, by the EC contract
HPRN-CT-2000-00148, by the INTAS contract 99-0-590 and by the MURST-COFIN
contract 2001-025492.

\begin{appendix}

\section{Notation}

Since we made many small notational changes with respect to
\cite{BerSam01}, where the domain wall solution was constructed and
discussed, it might be useful to collect the differences. The reason
for these differences is the discrepancy between the 3d supergravity
literature and holography literature (such as
\cite{BiFrSk01,BiFrSk01a}) and in this paper we stick close to the
latter.
The differences between  {\em old} (reference \cite{BerSam01}) and
{\em new} (this paper) notation are summarized in table~\ref{notation}.

\begin{table}[hn]
\centering
\begin{tabular}{|c||c|c|c|c|c|c|c|c|}\hline
{\rm old} & $q$          & ${\sf q}$ & $\phi^{\Si}$  &$\varphi $    
& $(x_i, y_i)$ & $g^2 V$ & $V_{\bf i}$  & $\kappa^2 $
\\
\hline
{\rm new} & $\sqrt{2}Q$  & $q$  &  $\Phi^{\Si}/\sqrt{2}$  &
$\phi^{\Si}$ 
& $(Z_1, Z_2)$ & $V$     & $V^{\rm tot}_{\bf i}$ 
& $\kappa $
\\ \hline
\end{tabular}
\caption{\small Notation: reference \cite{BerSam01} vs. this paper.}
\label{notation}
\end{table}

In addition, the two integration constants in the domain wall solution
\Ref{kink} are rescaled relative to \cite{BerSam01}, so that also the
expansions \Ref{BGexpansion} appear different; in both cases the
integration constants can be fixed by demanding $q_{(0){\rm B}}=1$ and
$g_{(0){\rm B}}=1$ in the domain wall solution.  Finally, the
signature in \cite{BerSam01} is Lorentzian $(+ - -)$ except for in
section 4 where it is Wick rotated to Riemannian signature, and here
it is Riemannian throughout.

\section{Analytics of the fluctuation equations}
\la{AHeun}

We have seen in the main text that the entire set of fluctuation
equations around the background \Ref{kink} may be reduced to equations
of the type
\ba
s \chi''_\alpha + (1\pls\alpha)\,\chi'_\alpha 
- P^2\, (2+s^3) \chi_\alpha &=& 0 \;, 
\la{odeA}
\ea
for $\alpha=0, -2$, and with $P=\sqrt{32/3}\,|p|$. This appendix is
devoted to a closer study of this differential equation and the
properties of its solutions. Let us recall that along the flow the
variable $s$ runs from $s=1$ at the AdS boundary (the UV) to
$s=\infty$ in the AdS interior. This implies that the two-point
correlation functions are encoded in the first coefficient of the
expansion at $s=1$
\ba
\chi_\alpha(s) &=& \chi_\alpha(1) \left(
1+ (s\mis 1) \,\Psi_{\alpha} + \dots \right)
\;,
\la{expt0}
\ea
of the solution $\chi_\alpha$ regular in the interior
$s\ra\infty$. The coefficient $\Psi_{\alpha}$ is uniquely defined as a
function of the parameter $P$ by the requirement that $\chi_\alpha$ is
regular as $s\ra\infty$. This procedure is entirely analogous to that
used in the analysis of the five-dimensional flows
\cite{BiFrSk01,Muck01}, where the corresponding differential equations
may be reduced to hypergeometric equations.  In those cases, however,
the regularity condition was imposed at a curvature singularity.

A first inspection shows that \Ref{odeA} has two singular points (zero
and infinity) with ``$s$-rank $\{1;3\}$'', in the language of
\cite{SlaLay00}.  This means that it may be obtained from a Fuchsian
differential equation with four regular singularities by making three
of these singularities coalesce at infinity.\footnote{For comparison,
the confluent hypergeometric equation has $s$-rank $\{1;2\}$.} This
equation is known as the biconfluent Heun equation~\cite{Ronv95}. With
the change of variables
\ba
\chi&=&e^{-\frac12 u^2} Y(u) \;,\qquad u~=~\sqrt{P}\, s\;,
\ea
equation \Ref{odeA} is mapped into a standard form
\ba
uY''+(1+\Ga-\Gb u-2u^2)\,Y'+\left(
(\Gg-2-\Ga)\,u-\ft12(\Gd+\Gb(1+\Ga))\right) Y &=& 0 \;,
\la{Heun}
\ea
with $(\Ga,\Gb,\Gg,\Gd)=\left(\Ga,0,0,4P^{3/2}\right)$. The solution
regular at $s=0$ is commonly denoted as $N(\Ga,\Gb,\Gg,\Gd;u)$. For
$\Gb=\Gd=0$, it reduces to a hypergeometric function; equation
\Ref{odeA} is a different (but also very special) case.

Despite considerable effort, see e.g.~\cite{Ronv95} and references
therein, the Heun equation is still far less understood than the
hypergeometric equation, which is the analogous fluctuation
equation in the previously studied flows.  For the purposes of
computing correlation functions, there is even an additional technical
complication here that comes from the fact that the relevant expansion
\Ref{expt0} (i.e.\ the UV boundary of the flow) is around a generic
regular point ($s=1$), rather than around a singular point as in the
higher-dimensional examples~\cite{BiFrSk01,Muck01}. Expanding around a
singular point considerably simplifies the resulting expressions; the
coefficient analogous to $\Psi$ in \Ref{expt0} for the case of the
hypergeometric equation is usually denoted as $\psi$ and is simply
expressed in terms of $\Gamma$-functions.  In all, it is a tall order
to solve \Ref{odeA}, but we find that we can extract the important
information analytically, and use numerics to check.

Following~\cite{Ronv95}, we denote by $H^+(\Ga,\Gb,\Gg,\Gd;u)$ the
unique solution of \Ref{Heun} that is regular as
$s\rightarrow\infty$. The relevant coefficient in \Ref{expt0} is then
given by
\ba
\Psi_\alpha(P) &=& -P + \frac{\partial}{\partial s}\, 
\log H^+\!\left(\Ga,0,0,4P^{3/2};\sqrt{P}\,s\right) \Big|_{s=1} \;.
\la{PsiHeun}
\ea
Analytic expressions for $H^+(\Ga,\Gb,\Gg,\Gd;u)$ may be constructed
along the lines of \cite{Exto91} as an infinite chain of sums over
Pochhammer symbols
$(x)_n=\Gamma(x+n)/\Gamma(x)$. \footnote{Unfortunately, the earlier
results of \cite{Exto88} which seemingly give $N(\Ga,0,0,\Gd)$ as a
series in hypergeometric functions are incorrect; the double sum in
eq.\ (2.8) in that paper does not, in fact, factor into (2.10); a
relation which is recursively used in the construction.}  Rather than
constructing these series, we will derive analytical results only for
the IR asymptotics in $P$, and use the numerical solution of
\Ref{PsiHeun} for other purposes.

To this end, we consider the following change of variables 
\ba
\chi&=& v^{-\alpha/4}\, Z(v) \;,\qquad v~=~\ft12 P\, s^2\;,
\ea
which transforms \Ref{odeA} into
\ba
\Delta_{\alpha/4} Z &\equiv &
v^2 Z'' + v Z' - \left(v^2 + \ft{\alpha^2}{16} \right) Z ~=~ 
P^{3/2}\, \frac{\sqrt{v}}{\sqrt{2}}\, Z  \;.
\la{serIR}
\ea
where prime now denotes a derivative with respect to $v$.
This equation may thus be resolved into a series of inhomogeneous
Bessel equations
\ba
Z &=& \sum_{r=0}^\infty P^{3r/2} \, 2^{-r/2} \, Z_r \;,\qquad 
\GD_{\alpha/4}\, Z_r ~=~ v^{1/2}\, Z_{r-1} \;.
\ea
which can successively be integrated in terms of 
Lommel functions,
demanding regularity in the interior $v=\infty$. Note that this
expansion is, in particular, 
compatible with the IR (small $P$) asymptotics. For instance, for
$\Ga=-2$, we explicitly find the first terms in this series as
\ba
Z_0 &=& \sqrt{2/\pi} \, K_{\frac12}(v) ~=~ v^{-\frac12} e^{-v}
\;,\qquad
Z_1 ~=~ \sqrt{2\pi} \,v^{-\frac12} e^{v} {\rm erfc}(\sqrt{2v})
\;,
\ea
with the complementary error function ${\rm erfc}(v)$. After some
computation, this gives rise to the small $P$ asymptotics
\ba
\Psi_{-2} &=& -\ft{\alpha}{2} + P\,\ft{\dd}{\dd v}\,\log Z
\Big|_{v=P/2} ~=~
 -P - P^2 + 2\sqrt{\pi} \,P^{5/2} - 4P^3 +  \dots  \;.
\la{IR2}
\ea
For $\Ga=0$, one finds similarly the lowest term $Z_0 = K_{0}(v)$, and
thus the small $p$ asymptotics
\ba
\Psi_0 &=&
\frac{2}{C+\log (P/4)} + \dots \;,
\la{IR0}
\ea
with Euler's constant $C$. As we have shown in the main text, the
two-point correlation functions of the inert scalars in the ${\bf 4}$
are proportional to the inverse of $\Psi_0$; the log-term then
describes the standard behavior of a dimension $\Delta=1$ operator in
the IR, cf.~table~\ref{specL0}. An analysis of the large $P$
asymptotics of the $\Psi_{\alpha}$ gives
\ba
\Psi_{-2} &=& -\sqrt{3}\,P + \ft 12 + \dots \;,
\qquad
\Psi_0 ~=~ -\sqrt{3}\,P - \ft 12 + \dots \;.
\la{UV}
\ea
To obtain a numerical expression for the important
ratio $\Psi_\Ga$, we can use the following simple prescription.
Consider the
differential equation \Ref{Heun} and numerically compute
two of its solutions with two different sets of initial 
conditions given at
$u=\sqrt{P}$ (the regular point $s=1$) as
\ba
Y_1(\sqrt{P}) &=& 0 \;,\qquad Y'_1(\sqrt{P}) ~=~ -\sqrt{P} \;,
\non
Y_2(\sqrt{P}) &=& 1 \;,\qquad Y'_2(\sqrt{P}) ~=~ 0  \;.
\ea
Since there exists a unique solution 
regular as $u \rightarrow \infty$ (called $H^+(u)$ above), the ratio
$Y_2/Y_1$ will tend to a constant in this limit, and 
this constant may be
determined numerically. From \Ref{PsiHeun}, we then 
have the relevant coefficient $\Psi_{\alpha}$ as
\ba
\Psi_\alpha &=& -P + \lim_{u\ra\infty}\,\frac{Y_2}{Y_1} \;.
\la{ratio}
\ea
The result for $\alpha=-2, 0$ is plotted in figures 3 and 4 together
with the first few terms of the exact asymptotics, given in
\Ref{IR2}--\Ref{UV}. Agreement in these regions is already quite good 
for including just a few terms.

\bigskip
\bigskip

\hspace*{-5mm}
\begin{minipage}[htbp]{75mm}
\resizebox{70mm}{!}{\includegraphics{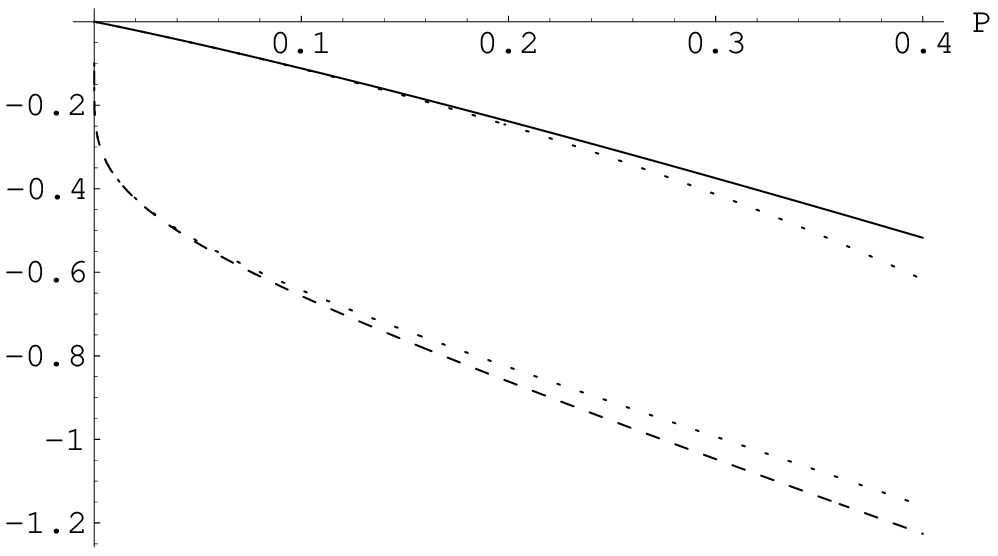}}
\medskip

\hspace*{5mm}
\begin{minipage}[htbp]{60mm}
{\small {Figure 3:} Small $P$ (IR) asymptotics of $\Psi_{-2}$
(straight), and $\Psi_0$ (dashed). The dotted lines correspond to the
first terms of the exact asymptotics \Ref{IR2}, \Ref{IR0}.}
\end{minipage}
\end{minipage}
\begin{minipage}[htbp]{80mm}
\resizebox{73mm}{!}{\includegraphics{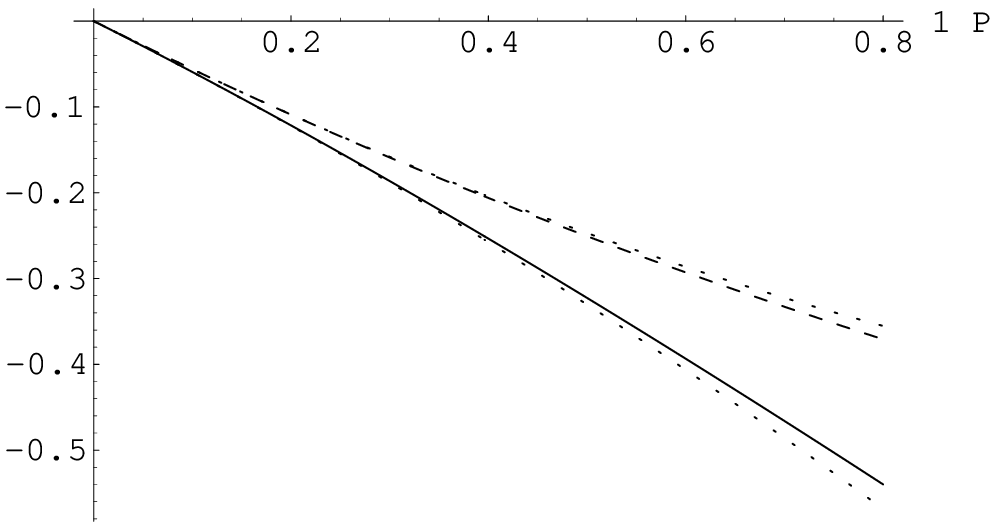}}
\medskip

\hspace*{8mm}
\begin{minipage}[htbp]{60mm}
{\small {Figure 4:} Large $P$ (UV) asymptotics of $1/\Psi_{-2}$
(straight), and $1/\Psi_0$ (dashed). The dotted lines correspond to
the first terms of the exact asymptotics \Ref{UV}. The horizontal axis
is $1/P$.}
\end{minipage}
\end{minipage}

\addtocounter{figure}{2}

\bigskip

\end{appendix}

%\bibliographystyle{Jopt2}
%\bibliography{bib}

\providecommand{\href}[2]{#2}\begingroup\raggedright\endgroup

\end{document}